\documentclass{ws-ijmpc}

\usepackage{subfigure}
\usepackage[super,numbers,sort&compress]{natbib}

\begin{document}

\markboth{J. Fang et al.}
{JAMMING TRANSITION OF POINT-TO-POINT TRAFFIC THROUGH COOPERATIVE MECHANISMS}

\catchline{}{}{}{}{}

\title{JAMMING TRANSITION OF POINT-TO-POINT TRAFFIC THROUGH COOPERATIVE MECHANISMS}

\author{JUN FANG\footnote{Corresponding author.}}

\address{Department of Computer Science and Technology\\
National Laboratory for Information Science and Technology\\
Tsinghua University, Beijing 100084, P. R. China\\
fangjun06@mails.tsinghua.edu.cn}

\author{ZHENG QIN}

\address{Department of Computer Science and Technology \\
Tsinghua University, Beijing 100084, P. R. China\\
qingzh@tsinghua.edu.cn}

\author{Xiqun Chen}
\address{Department of Civil Engineering\\
Tsinghua University, Beijing 100084, P. R. China\\
chenxq04@mails.tsinghua.edu.cn}

\author{Zhaohui Xu}

\address{Department of Computer Science and Technology \\
National Laboratory for Information Science and Technology\\
Tsinghua University, Beijing 100084, P. R. China\\
gogogofree@gmail.com
}

\maketitle


\begin{abstract}
We study the jamming transition of two-dimensional point-to-point traffic through cooperative mechanisms using computer simulation. We propose two decentralized cooperative mechanisms which are incorporated into the point-to-point traffic models: stepping aside (CM-SA) and choosing alternative routes (CM-CAR). Incorporating CM-SA is to prevent a type of ping-pong jumps from happening when two objects standing face-to-face want to move in opposite directions. Incorporating CM-CAR is to handle the conflict when more than one object competes for the same point in parallel update. We investigate and compare four models mainly from fundamental diagrams, jam patterns and the distribution of cooperation probability. It is found that although it decreases the average velocity a little, the CM-SA increases the critical density and the average flow. Despite increasing the average velocity, the CM-CAR decreases the average flow by creating substantially vacant areas inside jam clusters. We investigate the jam patterns of four models carefully and explain this result qualitatively. In addition, we discuss the advantage and applicability of decentralized cooperation modeling.
\end{abstract}

\ccode{PACS Nos.: 05.70.Fh; 05.90.+m; 89.40.-a; 89.90.+n}

\section{Introduction}
The traffic problems have attracted many scholars with different types of backgrounds and many traffic models have been developed in the physics literature \cite{Chowdhury2000, Helbing2001, Nagatani2002}. The two-dimensional traffic models are of considerable interest and have been used to study traffic problems related to the urban traffic \cite{Biham1992, Nagatani1993, Nagatani1995, Torok1996, Nagatani2009a, Benyoussef2003, Nagatani2008, Nagatani2009b, Chowdhury1999, Brockfeld2001, Shi2007, Moussa2007, In-nami2007, Huang2007, Fang2010}, pedestrian traffic \cite{Muramatsu1999, Muramatsu2000a, Muramatsu2000b, Jiang2007},  transportation of data packets on the Internet \cite{Ohira1998, Tretyakov1998},  path search of ant-like social insects \cite{Beckers1992, Chowdhury2002, John2004}, etc. In the urban traffic, there have been studies on traffic dynamic with two kinds of vehicles \cite{Biham1992},  anisotropic effect of cars in different travel directions \cite{Nagatani1993}, asymmetric exclusion model that takes into account jam-avoiding drive \cite{Nagatani1995},  green wave model \cite{Torok1996, Nagatani2009a},  anisotropy effect of the probabilities of changing motion directions \cite{Benyoussef2003}, vehicular traffic through a series of traffic lights \cite{Nagatani2008, Nagatani2009b}, combination of highway and urban traffic models \cite{Chowdhury1999, Brockfeld2001, Shi2007}, origin-destination effect of driver trips \cite{Moussa2007, In-nami2007, Huang2007, Fang2010}.

Mitchell and Durnota developed a two-dimensional cellular automaton (CA) which included adaptive route-changing behavior \cite{Mitchell1996}. In his model, every vehicle is associated with given origin-destination points and chooses the empty neighbor nearest to the destination as its next location. This rule is called the myopic or greedy approach. Moussa's CA model was based on Mitchell and Durnota's model  introducing the noise $\rho$ and the friction parameter $\mu$ \cite{Moussa2005}. Moussa's model updates in parallel order but Mitchell and Durnota's model updates in random sequence order. Maniccam studied four traffic models of point-to-point mobile objects and the first model was similar to Mitchell and Durnota's or Moussa's model \cite{Maniccam2004}. The other three models mainly differ in the O-D selection and the motion rule. Maniccam subsequently developed a four-way-biased random walker's model to study the effects of back step and update rule on the traffic jam \cite{Maniccam2005}.  In 2006, Maniccam proposed a new two-dimensional traffic model involving the adaptive decentralized congestion avoidance. In this model, each object doesn't choose the point nearest to the destination but the point with the lowest congestion level in its local region \cite{Maniccam2006}.

The cooperative mechanism exhibits explicitly or implicitly in the group of mobile objects with cooperation or intelligence, e.g. the ants or ant-like social insects \cite{Beckers1992, Chowdhury2002, John2004}, pedestrians \cite{Burstedde2001, Kirchner2004}, small robots. The point-to-point traffic models introduced above do not included any cooperative mechanism yet.

The purpose of this paper is to study the jamming transition of two-dimensional point-to-point traffic through cooperative mechanisms using computer simulation. It is supposed that the mobile objects have basic adaptability or intelligence and they are able to cooperate with each other for some goals or under some principles. We propose two cooperative mechanisms which are incorporated into the point-to-point traffic models: stepping aside (CM-SA) and choosing alternative routes (CM-CAR). We are interesting whether the two cooperative mechanisms could increase the critical density, average velocity and average flow of the system, or make it worse. The potential applications of our models with cooperative mechanisms include modeling and simulation in mechanical, biological, economical, network and sociological systems.

The paper is organized as follows. Section \ref{model} describes four traffic models. For model 1, mobile objects travel between randomly chosen origins and destinations without any cooperative mechanism. Based upon model 1, the model 2 and 3 incorporate CM-SA and CM-CAR, respectively. The model 4 combines the model 2 with model 3 to test the comprehensive effects of the two cooperative mechanisms. Section \ref{simulation} presents and compares our numerical results of the four models. Section \ref{conclusions} gives the summary and discussion of our work.

\section{Model} \label{model}

\subsection{Model 1: Maniccam's model}
The Model 1 is a classical point-to-point traffic model, which is the same as Maniccam's model \cite{Maniccam2004}. Mobile objects move from point to point on a square lattice with the closed boundary. When an object arrives at one of lattice boundaries, it can either back step or side step but not cross the boundary. The number of objects on the lattice remains constant. The excluded-volume effect is taken into account, that is, each point can hold one object at most.

Initially, each object is associated with given origin-destination points (OD). Each pair of OD is chosen randomly on the lattice and the origin and destination must be different. The origins of all objects must be different but their destinations are allowed to overlap. Then each object travels from its origin to its destination according to some traffic rules and above-mentioned cooperative mechanisms, which are described in the next subsections. After arriving at its destination, each object uses the destination as its new origin and randomly chooses another point as its new destination. After that they continue to travel from point to point unless the simulation time has reached the upper limit.

\begin{figure}[!ht]
\subfigure[]{
\label{model1:a} 
\begin{minipage}[b]{0.5\textwidth}
\centering
\includegraphics[width=1.8in]{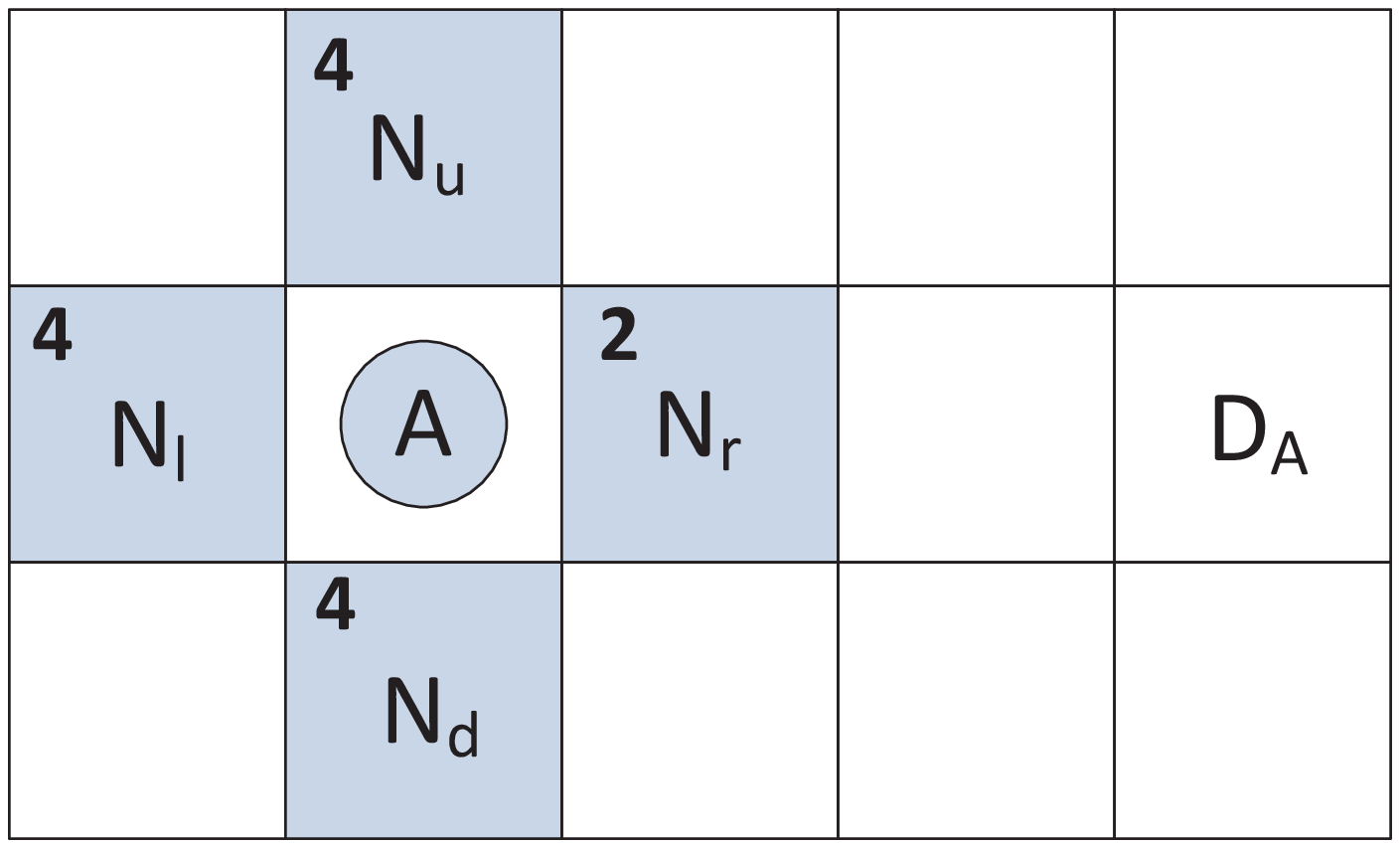}
\end{minipage}}%
\subfigure[]{
\label{model1:b} 
\begin{minipage}[b]{0.5\textwidth}
\centering
\includegraphics[width=1.8in]{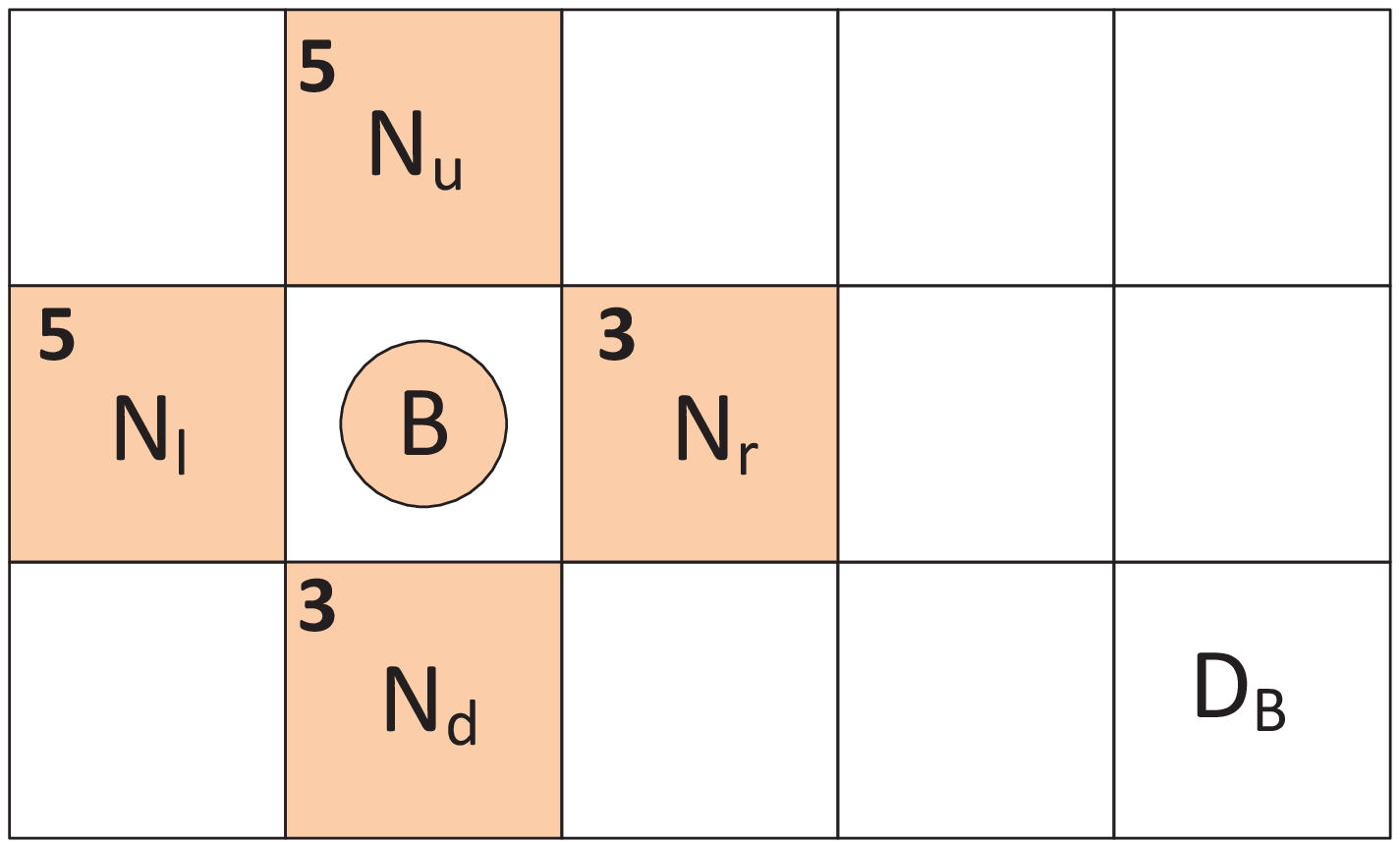}
\end{minipage}}
\caption{The neighborhood of an object and the distance from its each neighbor to its destination. These are two typical cases and the other cases are similar with the both: (a) The current location of the object and its destination are horizontally aligned; (b) The current location of the object and its destination are diagonally aligned.
}
\label{model1} 
\end{figure}

Fig.~\ref{model1} shows the neighborhood of an object and the distance from its each neighbor to its destination. The destination of object A and B is denoted by $\textrm{D}_{\textrm{A}}$ and $\textrm{D}_{\textrm{B}}$, respectively.  The digit in the upper left corner of each neighbor represents the distance from its current location to the destination of central object. The object selects the neighbor to move according to the myopic or greedy approach, i.e. the closest to the destination of the central object to move, unless the point has been occupied. When having more than one candidate, the object will choose one candidate randomly as its location next time. The object will move as long as it has a vacant neighbor, even though it goes far away from its destination next time. Because of the parallel rules, some points may be occupied by more than one object. In this case, the point will randomly choose a competitor as its occupant and leave the others waiting until next time.

\subsection{Model 2: cooperative mechanism of stepping aside (CM-SA)}

\begin{figure}[!htb]
\centering
\includegraphics[width=0.7\textwidth]{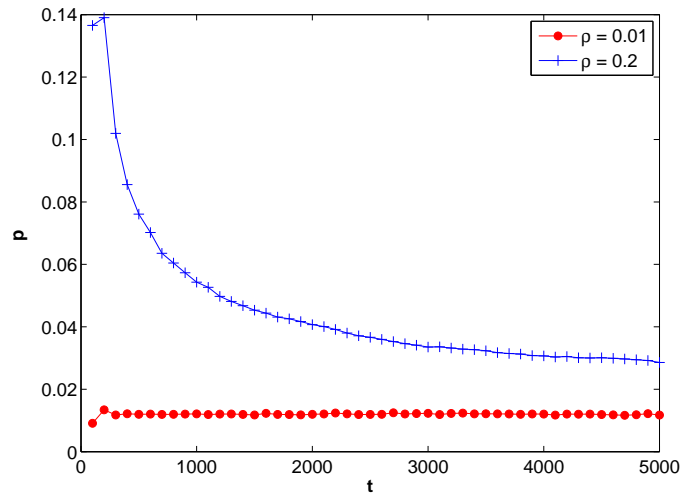}
\caption{Plots of the probability $p$ of ping-pong jump against time step $t$ for model 1 when system size $w=400$ and density $\rho=0.01, 0.2$. The probability $p(t)$ at time step $t$ is defined as the number of objects that undergo the ping-pong jump at time step $t$ divided by the total number of objects. Each probability curve is averaged over 10 different randomly initial conditions.}
\label{pp}
\end{figure}

However, the model 1 can generate the ping-pong jump, which delays objects arriving at their destinations and slows down the average flow of the system. To highlight the necessity to incorporate the cooperation mechanism, we give a typical plots of the probability $p$ of ping-pong jump against time step $t$ for model 1 in Figure~\ref{pp}.

We mainly discuss a type of ping-pong jump when two objects standing face-to-face want to move in opposite directions. This is because preventing this ping-pong jump from happening needs a pair of objects to cooperate with each other. Fig.~\ref{md1} show why the ping-pong jump could happen. According to the rule of model 1, when two objects meet together and step to the same side, they will do the ping-ping jump. We develop the first model by incorporating a cooperative mechanism called ``stepping aside'' (CM-SA), which is described as follows:
\begin{figure}[!ht]
\centerline{\psfig{file=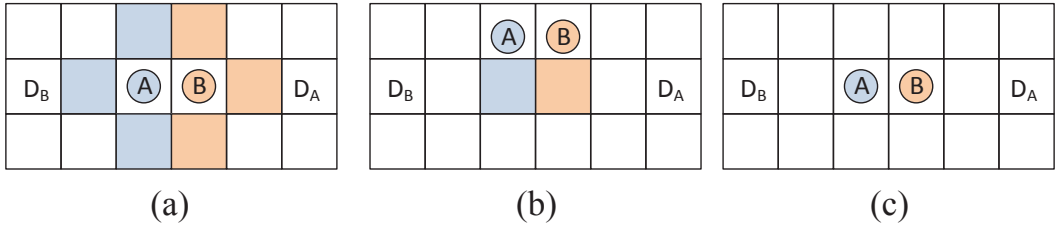, width=14cm}}
\caption{Illustration of a type of ping-pong jump when two objects standing face-to-face want to move in opposite directions. }
\label{md1}
\end{figure}

\begin{figure}[!ht]
\centerline{\psfig{file=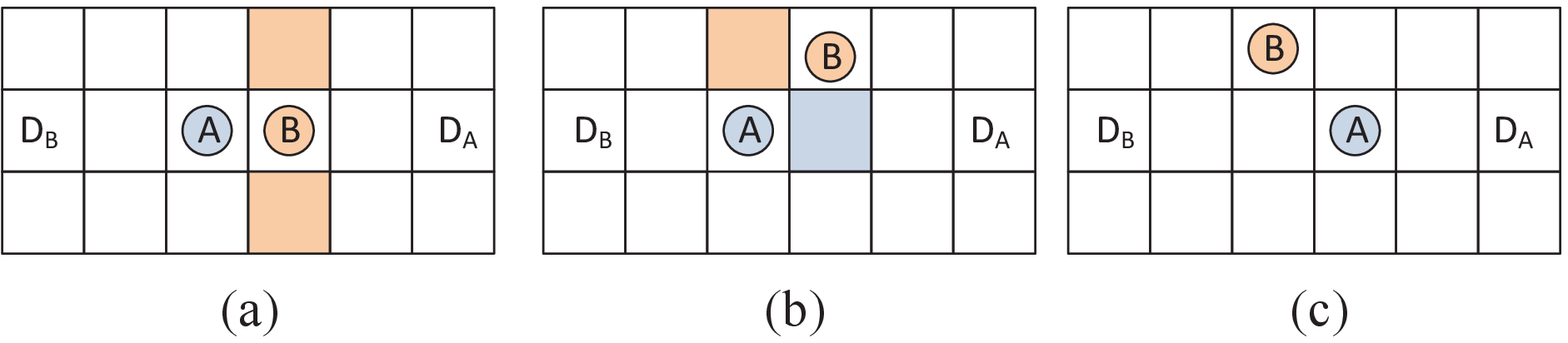, width=12cm}}
\caption{Illustration of the cooperative mechanism of stepping aside in model 2.}
\label{m2r1}
\end{figure}
\begin{figure}[!ht]
\centerline{\psfig{file=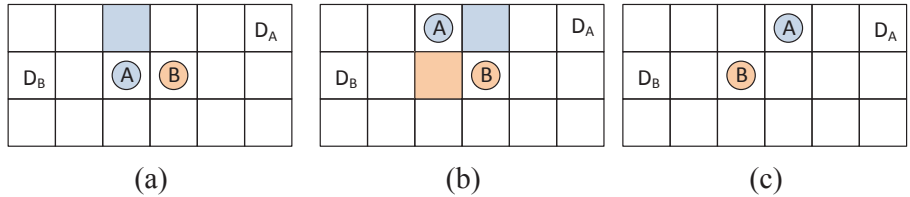, width=12cm}}
\caption{Illustration of the cooperative mechanism of stepping aside in model 2 as another example. }
\label{m2r2}
\end{figure}

(1) Three conditions for CM-SA must be satisfied. (a) A pair of objects stands face-to-face horizontally or vertically; (b) At least for one of the pair of objects, its current point and destination is in a perpendicular or vertical line; (c) At least one of the pair of objects has a vacant point on the side.

(2) One stands still and the other steps aside. For both objects in a pair, if their current point and destination are horizontally or vertically aligned, one object step aside and the other stands still. The avoidance behavior of one object gives way to the other directly facing its destination. Which one stepping aside is  chosen randomly.  If both sides of the moving object are vacant, it will step aside randomly,which is shown in Fig.~\ref{m2r1}(b). If the current point and destination of an object isn't in a perpendicular or vertical line, it will moves aside and the other stands still. Through an object stepping aside, both of them can directly face their  destinations, which is illustrated in Fig.~\ref{m2r2}(b).

(3) Priority of the cooperative mechanism. The cooperators in step (2) have some priority in occupying the lateral  vacant positions. The objects that take in the cooperations have higher priority than the others that don't do it. But two objects that both take in cooperations have the same priority. This setting is to increase the probability of success for the cooperation.

\subsection{Model 3: cooperative mechanism of choosing alternative routes (CM-CAR)}

\begin{figure}[!htb]
\centering
\includegraphics[width=0.7\textwidth]{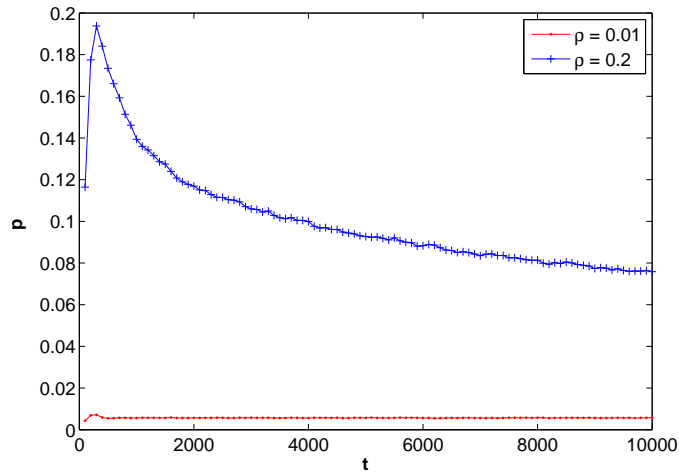}
\caption{Plots of the probability $p$ of a point competed by multiple objects against time step $t$ for model 1 when system size $w=400$ and density $\rho=0.01, 0.2$. The probability $p(t)$ at time step $t$ is defined as the number of points that multiple objects want to occupy at time step $t$ divided by the total number of points occupied by objects (equaling the total number of objects). Each probability curve is averaged over 10 different randomly initial conditions.}
\label{jz}
\end{figure}

According to the rules of model 1, when a point has more than one competitor, it will randomly choose a competitor and leave the others waiting until next time. This conflict resolution wastes the optional points of other objects if they have more than one candidate. And the conflict resolution decreases the average velocity of the system and may slow down the average flow of the system. To highlight the necessity to incorporate the cooperation mechanism, we give a typical plots of the probability $p$ of a point competed by multiple objects against time step $t$ for model 1 in Figure~\ref{jz}.

We develop the first model by incorporating a cooperative mechanism called ``choosing alternative routes'' (CM-CAR). We modify the conflict resolution to let competitors choose their alternative routes in turn according to the number of candidates. The cooperative mechanism in model 3 is described as follows:

\begin{figure}[]
\centerline{\psfig{file=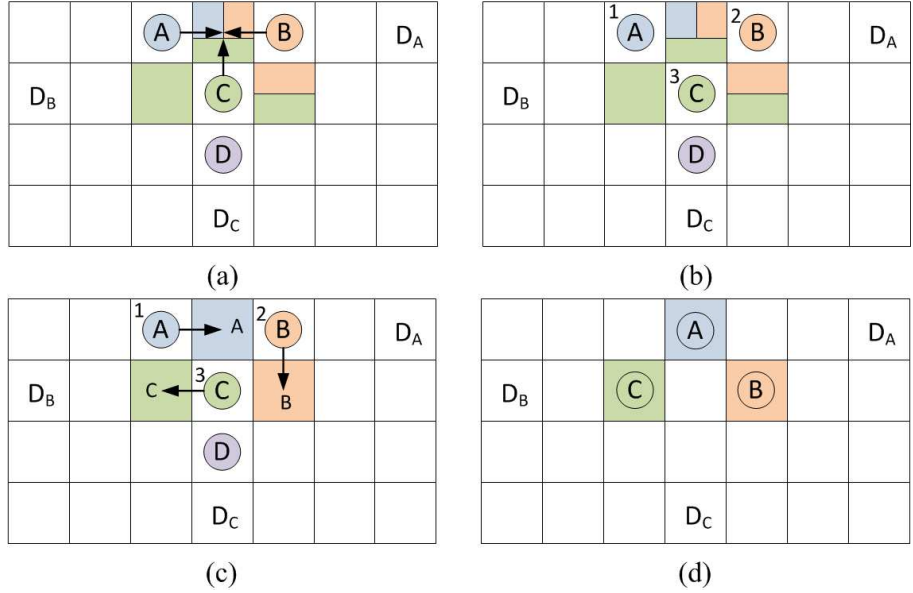, width=12cm}}
\caption{Illustration of cooperative mechanism of choosing alternative routes in model 3. The boxes colored with blue, pink, green correspond to the candidates of object A, B, C, respectively. If some boxes include multiple colors, they are as the candidates of multiple objects. }
\label{m31}
\end{figure}

(1) Sort competitors by the number of candidates. The object with the least candidates is allow to choose firstly. Then competitors choose their alternative routes in turn according to the number of candidates, which is illustrated in Fig.~\ref{m31}(b). If some objects have the same number of candidates, they are sorted randomly.

(2) Move one by one and mark occupied points. When sorting has completed, the competitors will choose their alternative routes one by one (see Fig.~\ref{m31}(c)). If a point has been occupied, regardless of by competitors or by other objects, it will be marked and not be chosen by any successor. The successor has to choose alternative routes if its original route is blocked. If all candidates of an object are occupied, it can only wait until next time.

\subsection{Model 4: the combined model}
The model 4 combines model 2 with model 3 to test the comprehensive effect of the two types of cooperative mechanisms. The model 4 is described as follows:

(1) The objects that can take part in the cooperation of stepping aside (CM-SA) are filtered  in pairs. Then the objects satisfying cooperative conditions move according to the definition of cooperative mechanism in model 2.

(2) The objects that can't take part in the cooperation in step (1) move according to the traffic rules of model 1.

(3) All points on the lattice are scanned to find which points will be occupied by more than one object. Then these competitors move according to the definition of cooperative mechanism (CM-CAR) in model 3. The competitors that have taken part in  cooperation (CM-SA) in model 2 have the priority over the others in occupying the vacant points. And they are put in the head in the order of choice.

\section{Simulation and results} \label{simulation}

The system size and the objects density are denoted by $w$ and $\rho$, respectively. So the number of objects on the lattice is $N=w^{2}\ast\rho$. We investigate four typical system sizes $w=\{50,100,200,400\}$.  The simulations at every density are repeated 10 times independently with different randomly initial conditions and each simulation is run for 10,000 time steps. At time step $t$, the average velocity $v(t)$ of the system is defined as the ratio of moving objects to all objects on the lattice. The ensemble average velocity $\langle v\rangle$ of the system is defined as the average of $v(t)$ over  $5,001\leq t \leq10000$ time steps. At time step $t$, the average flow $f(t)$ of the system is defined as the number of objects that arrive at their destinations at time step $t$. The ensemble average flow $\langle f\rangle$ of the system is defined as the average of $f(t)$ over $5,001\leq t \leq10000$ time steps.

\begin{figure}[]
\subfigure[]{
\label{v:a}
\begin{minipage}[]{0.5\textwidth}
\centering
\includegraphics[width=2.7in]{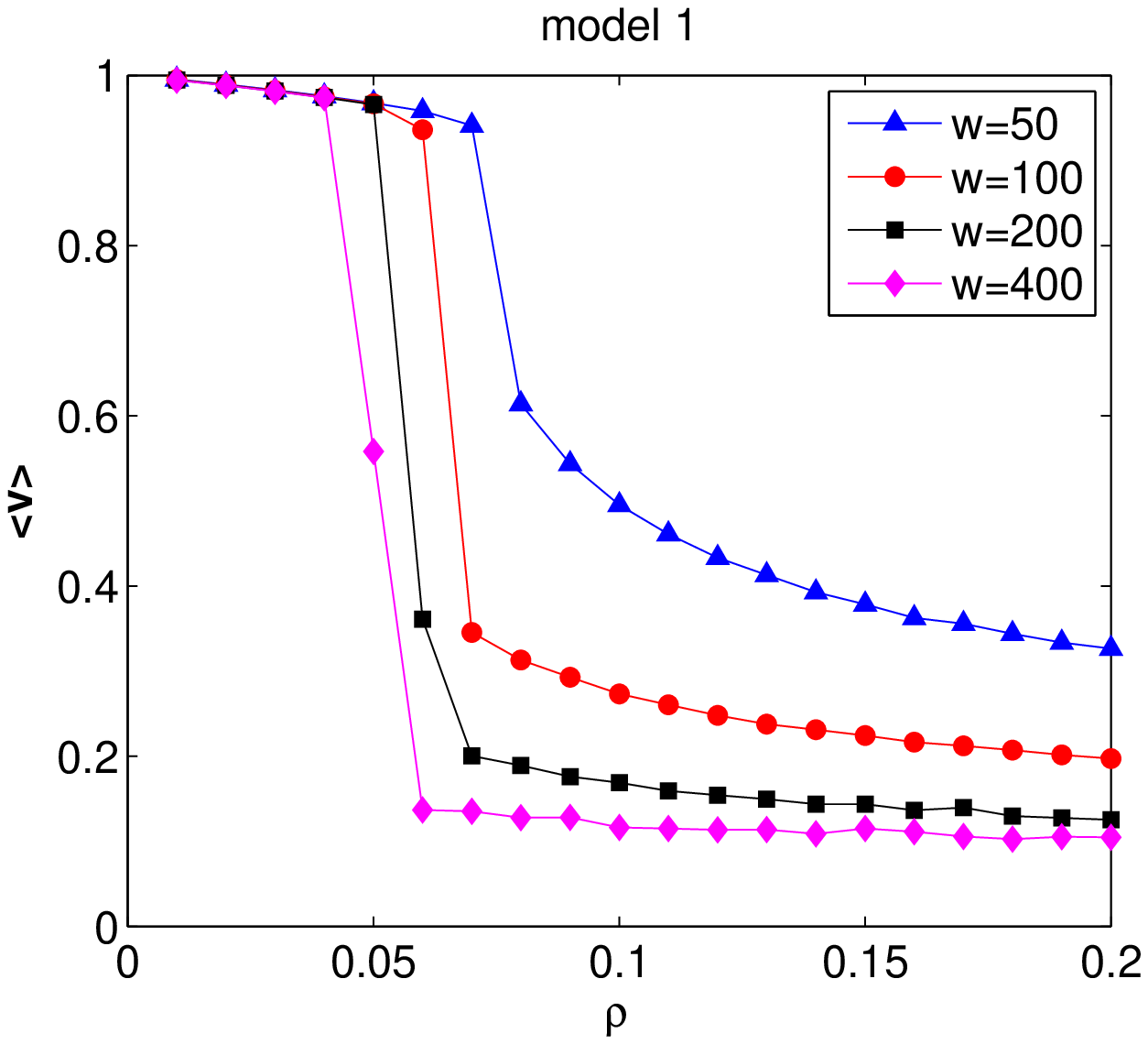}
\end{minipage}}%
\subfigure[]{
\label{v:b}
\begin{minipage}[]{0.5\textwidth}
\centering
\includegraphics[width=2.7in]{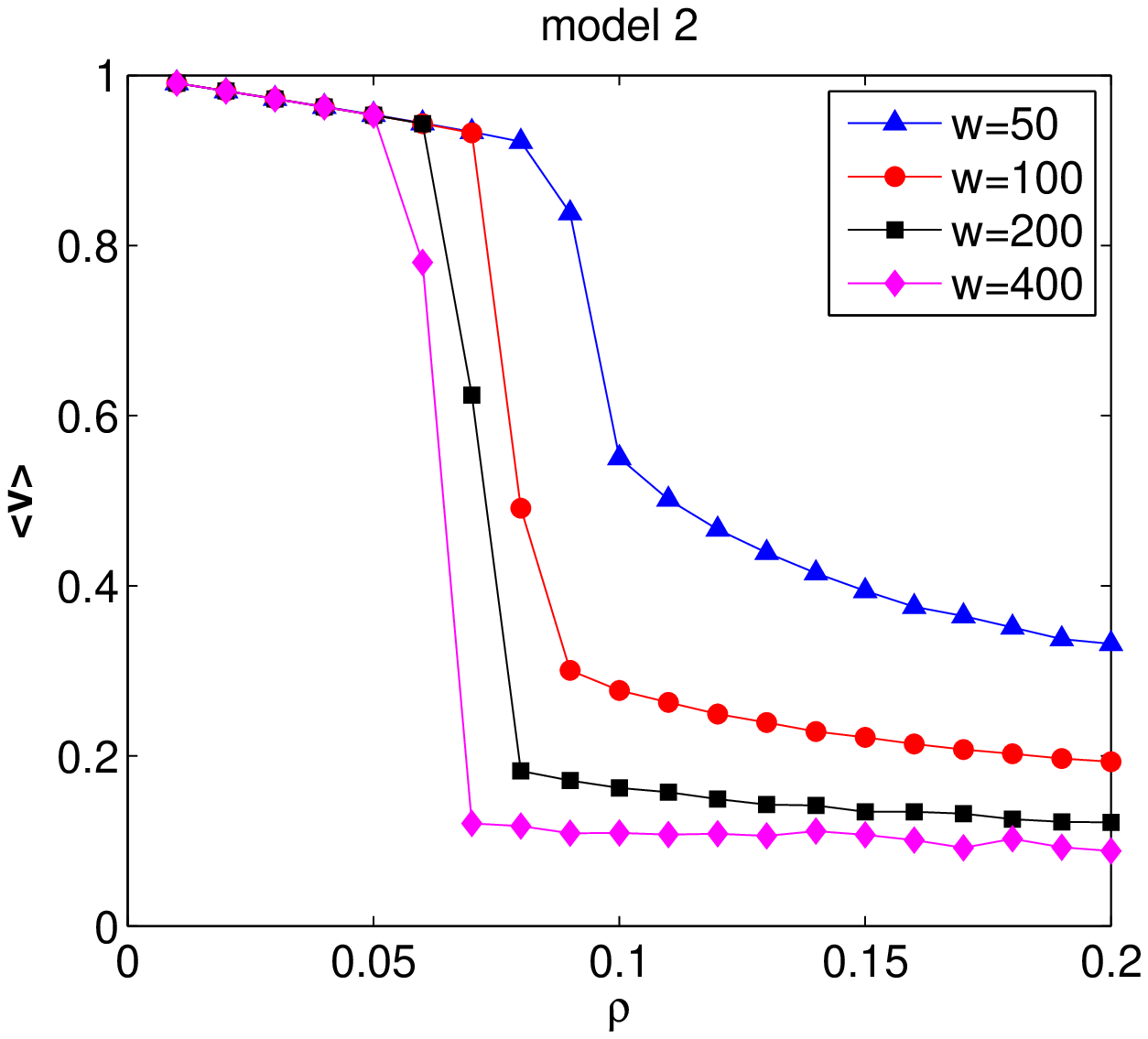}
\end{minipage}} \\
\subfigure[]{
\label{v:c}
\begin{minipage}[]{0.5\textwidth}
\centering
\includegraphics[width=2.7in]{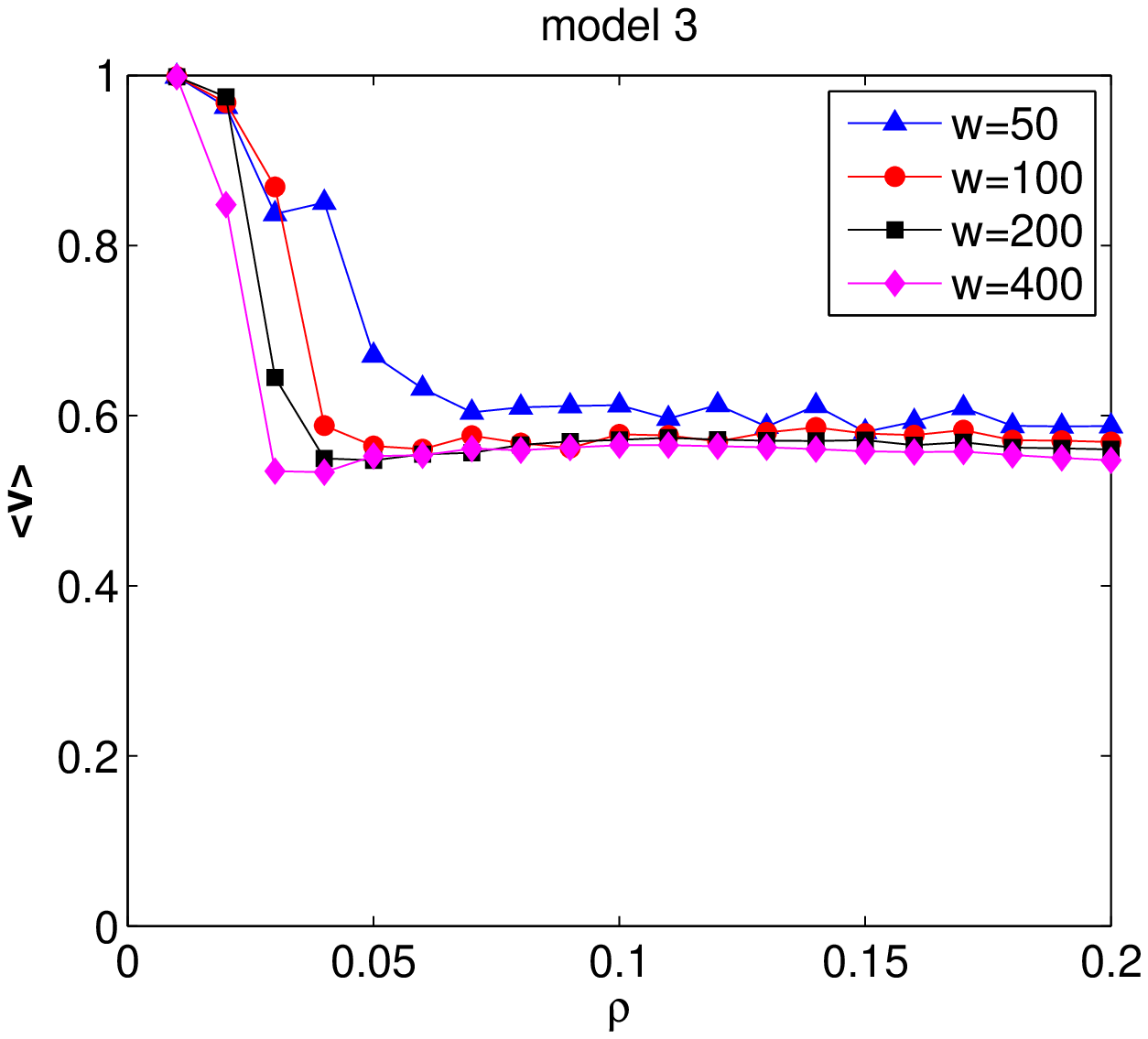}
\end{minipage}}%
\subfigure[]{
\label{v:d}
\begin{minipage}[]{0.5\textwidth}
\centering
\includegraphics[width=2.7in]{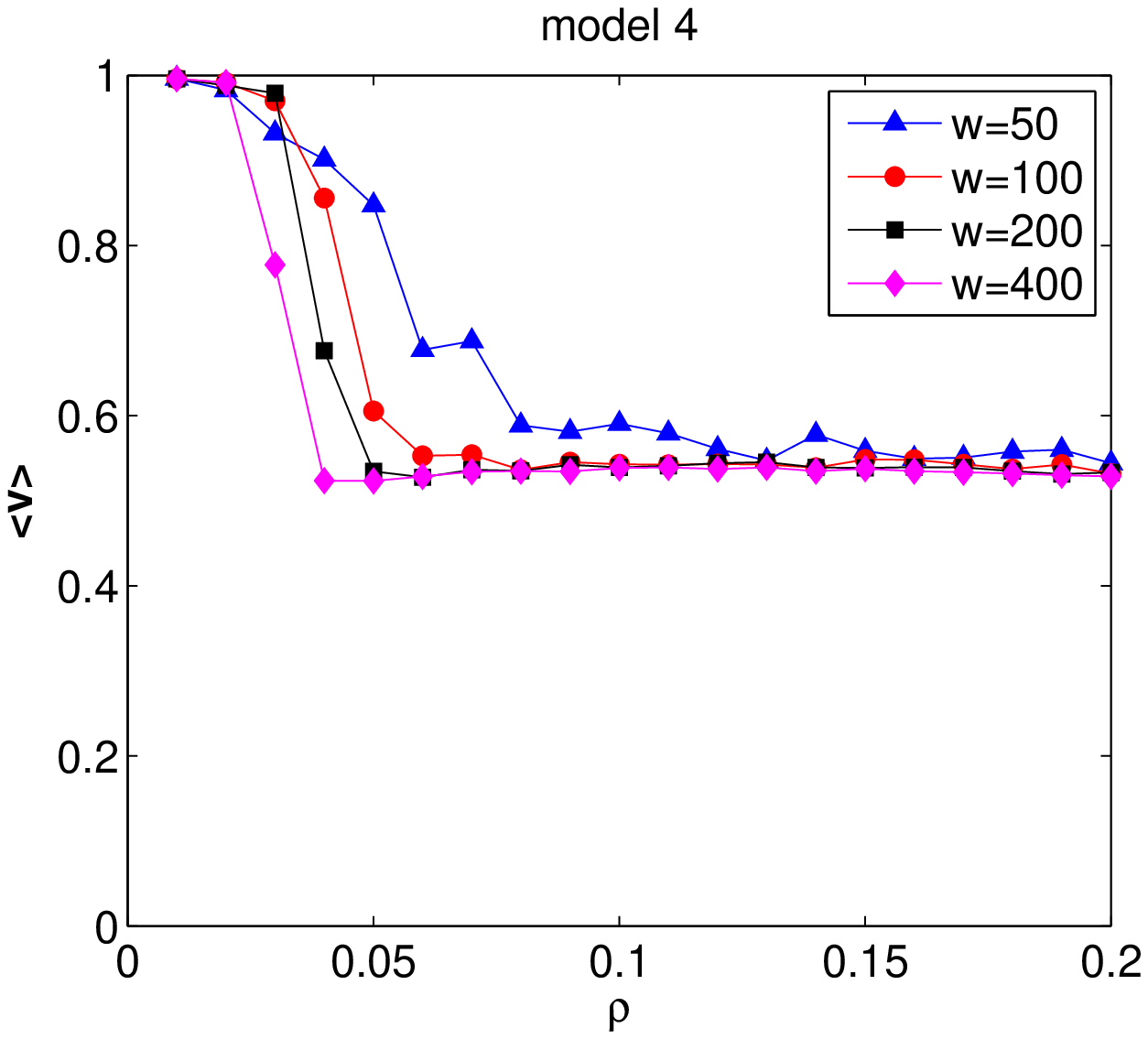}
\end{minipage}}
\caption{Diagrams of the average velocity $\langle v\rangle$ against density $\rho$ of four models for system size $w=50, 100, 200, 400$. The fundamental diagrams are plotted with density increasing with the step length of 0.01.
}
\label{v}
\end{figure}

For all velocity diagrams in Fig.~\ref{v}, there is a transitional region where the system undergoes a continuous transition from a free phase to a jam phase at the critical density $\rho_{c}$. Comparing different system sizes in the same model, we find $\rho_{c}$ keeps decreasing as $w$ increases, which coincides with the results of Ref.~\refcite{Biham1992}. For model 3 and 4, the system gets into a partial jam after the critical density with average velocity maintains around 0.55, until the density has reached about 0.6. After that, the average velocity goes down smoothly to zero (not shown in this paper). We compare the average velocity $\langle v\rangle$ of four models at density $\rho=0.2$  for the four system size. From the lowest to the highest value in order, they are model 2, model 1, model 4 and model 3. In Table~\ref{tab_density}, we obtain numerical results for critical density $\rho_{c}$ of four models. From the lowest to the highest value in order, they are model 3, model 4, model 1 and model 2. The cooperative mechanism of stepping aside (CM-SA) decreases the average velocity but increases the critical density. The cooperative mechanism of choosing alternative route (CM-CAR) decreases the critical density but increases the average velocity.

\begin{table}[]
\tbl{The critical density $\rho_{c}$ of model 1 - model 4 for four system sizes. The critical density is defined at the center of transitional region.}
{\begin{tabular}{@{}ccccc@{}} \toprule
w & model 1 & model 2 & model 3 & model 4 \\\colrule
\hphantom{0}50  & 0.08 & 0.09 & 0.05 & 0.06 \\
100 & 0.07 & 0.08 & 0.03 & 0.05 \\
200 & 0.06 & 0.07 & 0.03 & 0.04 \\
400 & 0.05 & 0.06 & 0.02 & 0.03 \\ \botrule
\end{tabular}
\label{tab_density}}
\end{table}

\begin{figure}[]
\subfigure[]{
\label{f:a}
\begin{minipage}[]{0.5\textwidth}
\centering
\includegraphics[width=2.7in]{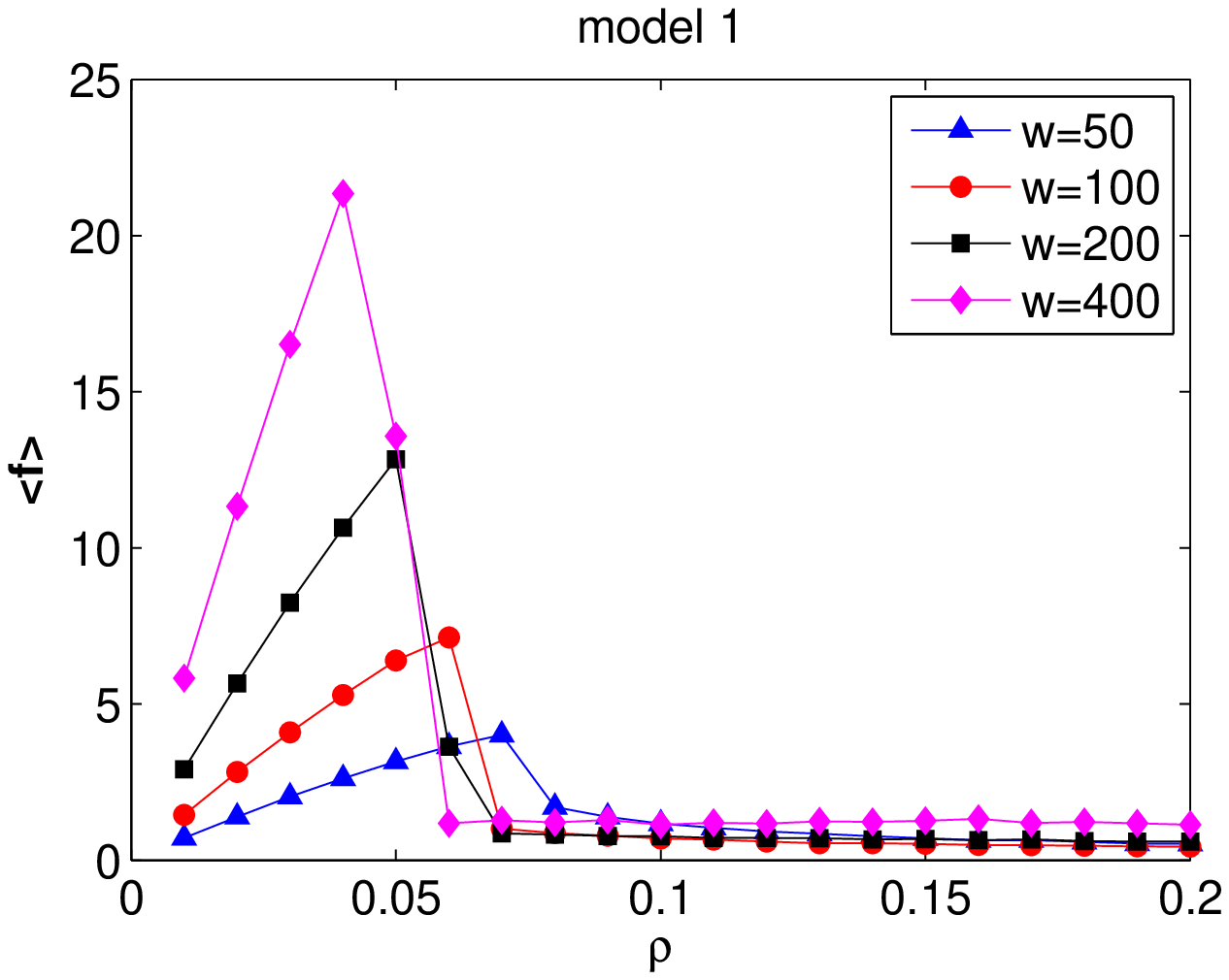}
\end{minipage}}%
\subfigure[]{
\label{f:b}
\begin{minipage}[]{0.5\textwidth}
\centering
\includegraphics[width=2.7in]{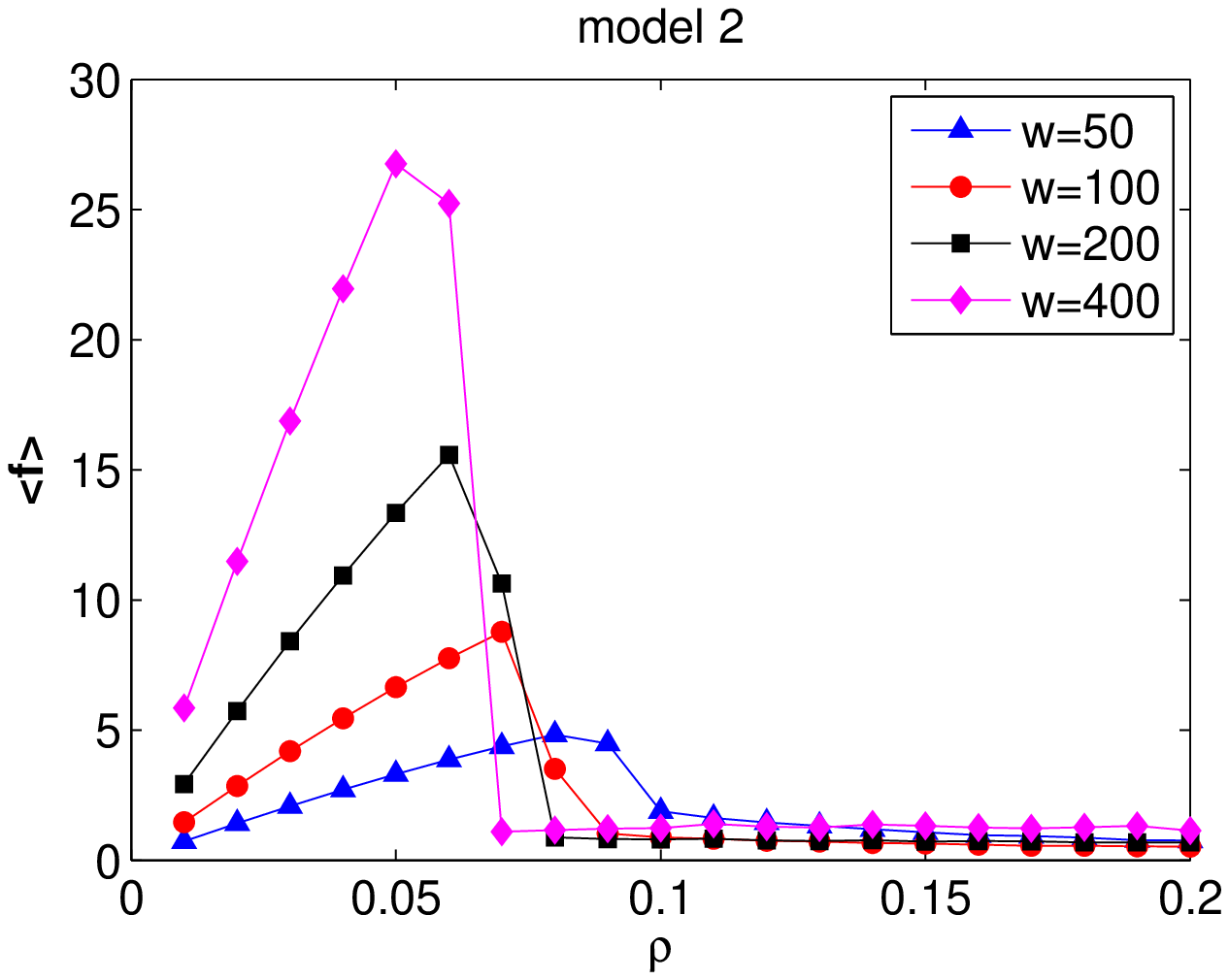}
\end{minipage}} \\
\subfigure[]{
\label{f:c}
\begin{minipage}[]{0.5\textwidth}
\centering
\includegraphics[width=2.7in]{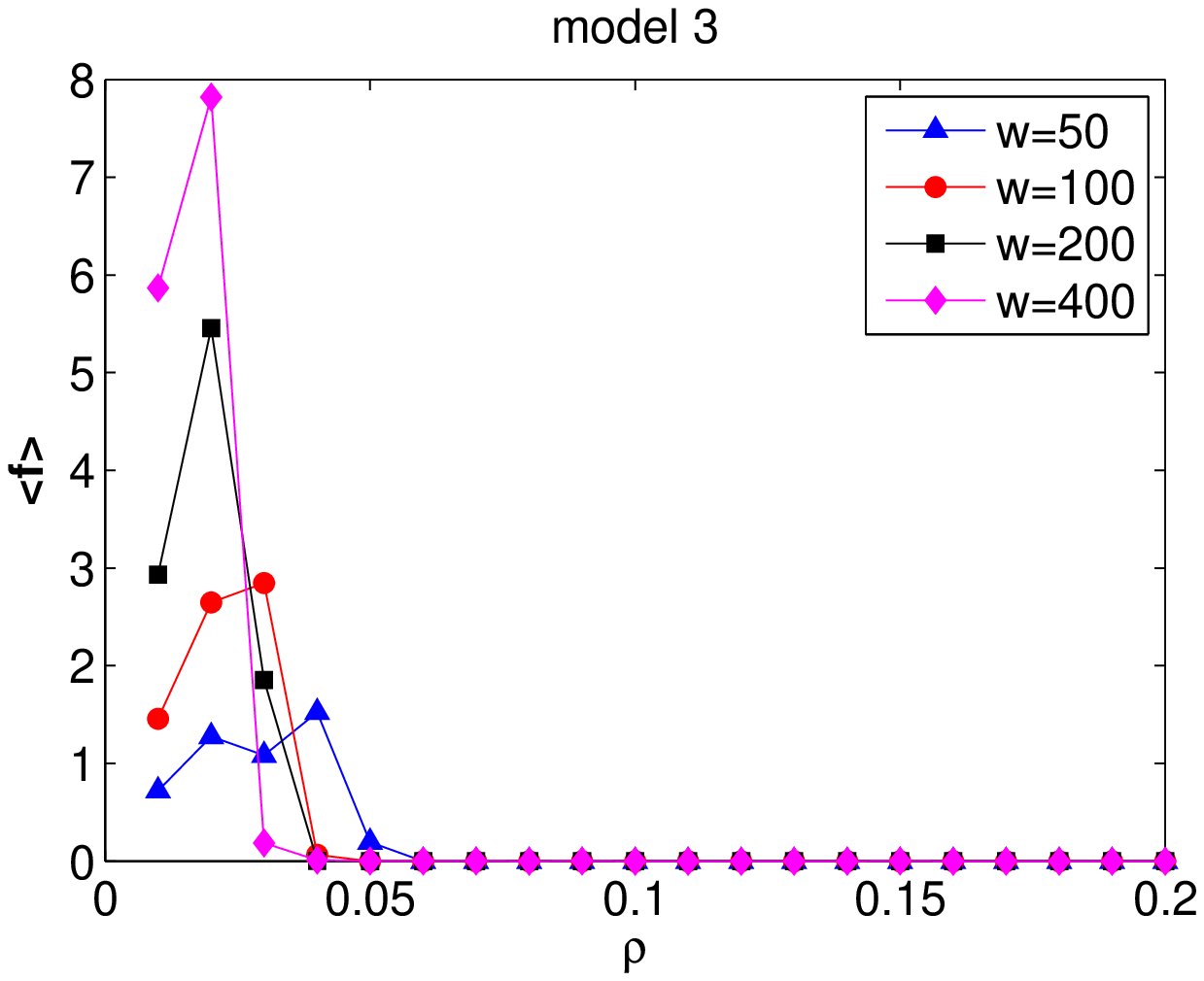}
\end{minipage}}%
\subfigure[]{
\label{f:d}
\begin{minipage}[]{0.5\textwidth}
\centering
\includegraphics[width=2.7in]{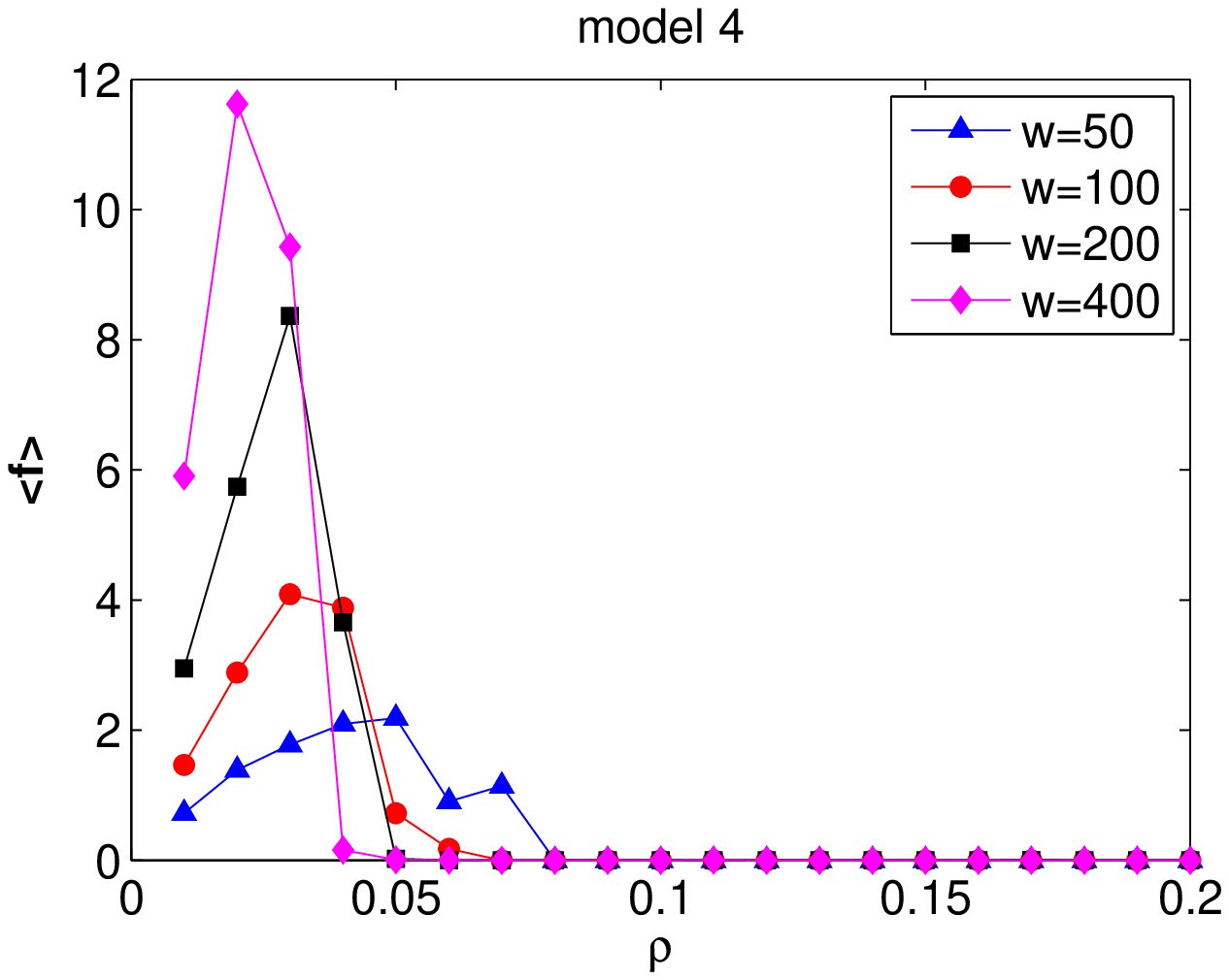}
\end{minipage}}
\caption{Diagrams of the average flow $\langle f\rangle$ against density $\rho$ of four models for system size $w=50, 100, 200, 400$. The fundamental diagrams are plotted with density increasing with the step length of 0.01.
}
\label{f}
\end{figure}

In Fig.~\ref{f}, With increasing the system size $w$, the max flow increases and the critical density decreases. It is noted that despite higher average velocity, the average flow of model 3 and 4 is lower than model 1 and 2.  This result indicates that fewer objects can arrive at their destination for model 3 and 4 than model 1 and 2. Why this would happen will be illustrated in the following.

Because each object does not always move in the shortest path, the travel route may have side steps, back steps, detours, loops and waits. The flow is contributed by the average velocity and the ratio of advanced movement simultaneously, not only the average velocity. For model 3 and model 4, although the velocity is increased obviously, the ratio of advanced movement does not be increased with the increase of the velocity. The increase in the amount of advance is canceled by the increase in the amount of detour. How to increase the ratio of advanced movement in model 3 and model 4 is worth further research.
\begin{figure}[!htb]
\centering
\subfigure[t=989]{
\label{model4_01}
\includegraphics[width=0.45\textwidth]{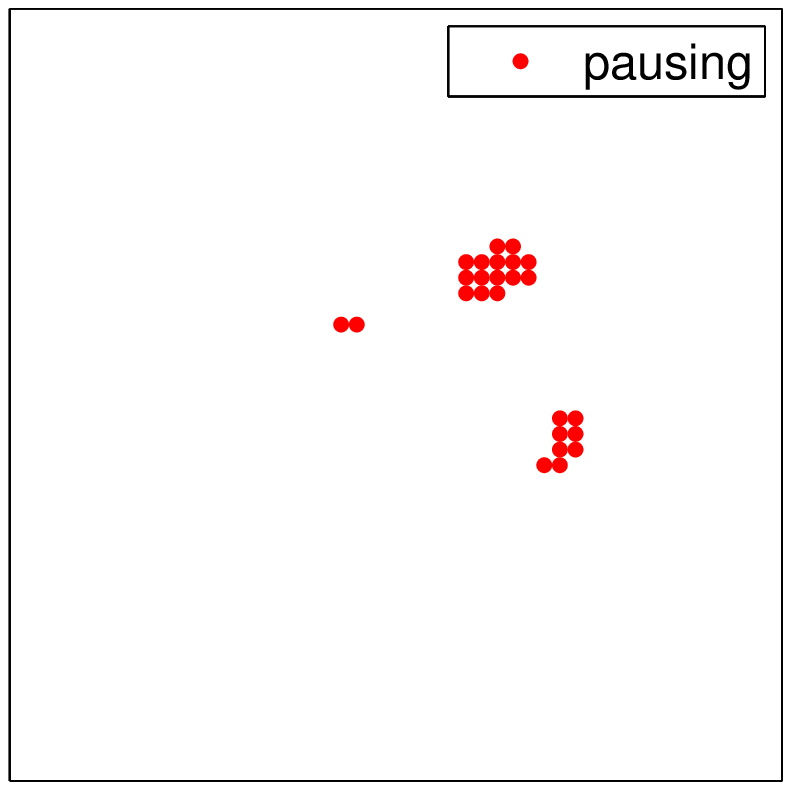}
}
\hfill
\subfigure[t=10,000]{
\label{model1_07}
\includegraphics[width=0.45\textwidth]{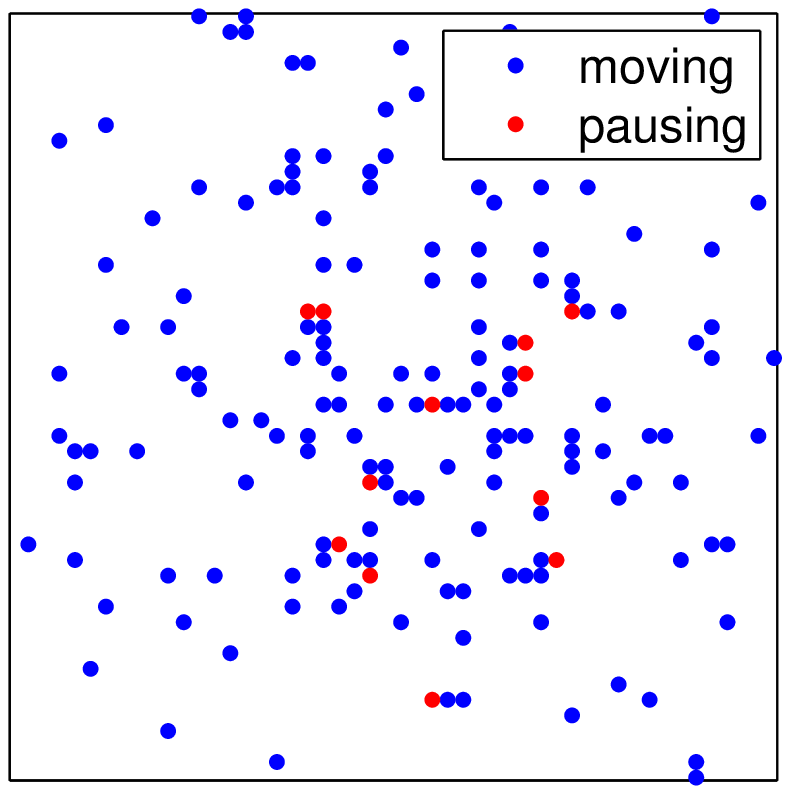}
}
\caption{The typical traffic configurations of two models at $L=50$: (a) the model without detour (the fourth model in Ref.~30) at $\rho=0.01$, (b) the model with detour (model 1) at $\rho=0.07$.}
\label{fourth_model}
\end{figure}

Is the detour useless when the objects move sideways or backwards and go far away from their destinations? Actually, the detour effectively reduces traffic jam and improves the critical density, especially when there are four moving directions and no separated lanes for traffic in opposite directions. In order to verify the effectiveness of detour, we reproduce the experiments of the fourth model in Ref.~\refcite{Maniccam2004} and show a typical traffic configuration on the lattice in Figure~\ref{model4_01}. The fourth model in Ref.~\refcite{Maniccam2004} is similar to the model 1 except that each object only travels in the shortest path without detour. If there are obstacles, an objects will wait until it can move again. In Figure~\ref{model4_01}, all objects are trapped into a complete jam after $t=989$ even $\rho=0.01$. While for the traffic in Figure~\ref{model1_07}, the system is still at free phase after $t=10,000$ at $\rho=0.07$. If the objects have to move ahead without detour, the objects coming from four directions will form a gridlock easily. In this gridlock, everyone is waiting for others to make way for itself, and as a result, on one can move again.
\begin{figure}[]
\subfigure[]{
\label{j_2_10000:a}
\begin{minipage}[]{0.5\textwidth}
\centering
\includegraphics[width=2.5in]{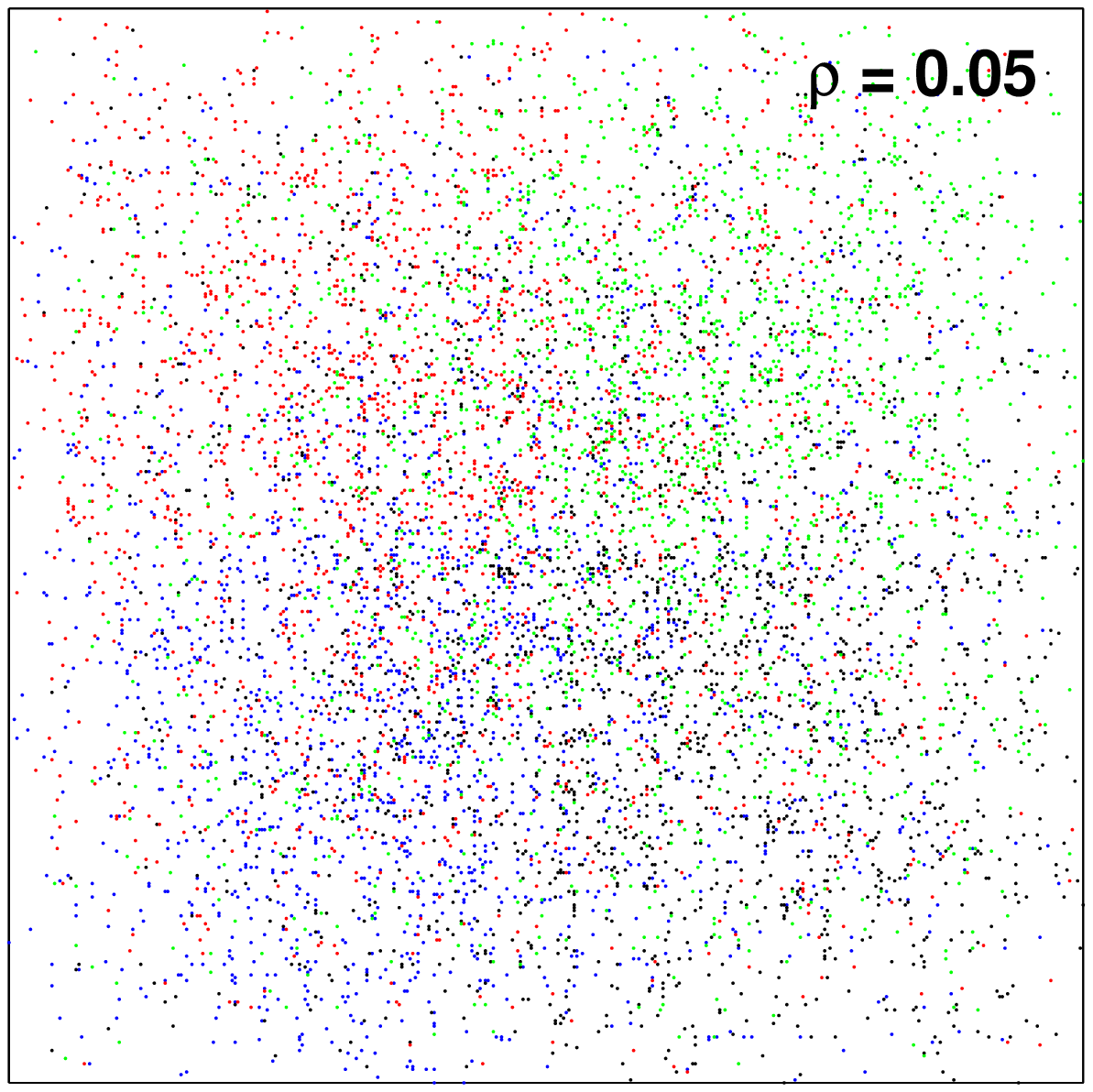}
\end{minipage}}%
\subfigure[]{
\label{j_2_10000:b}
\begin{minipage}[]{0.5\textwidth}
\centering
\includegraphics[width=2.5in]{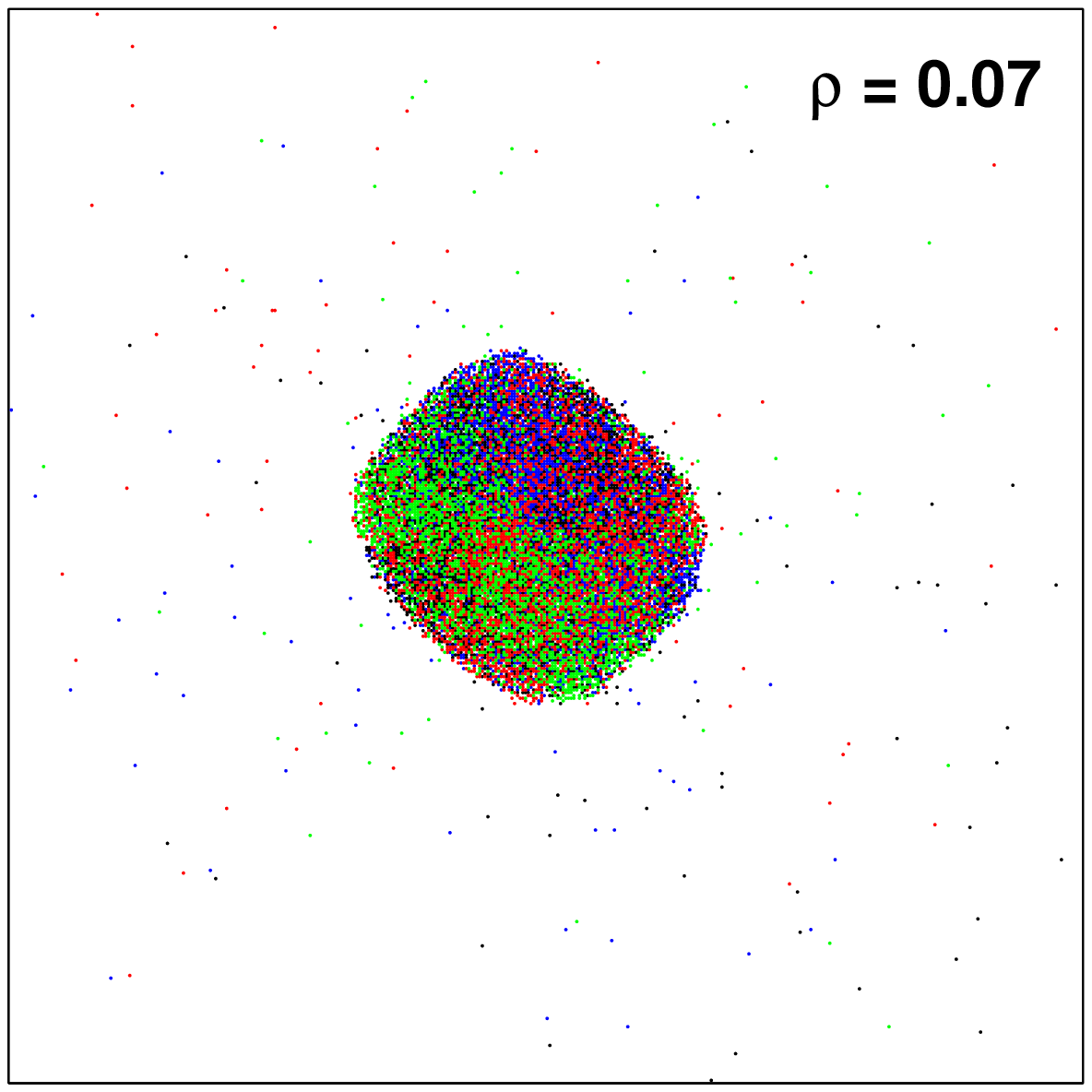}
\end{minipage}} \\
\subfigure[]{
\label{j_2_10000:c}
\begin{minipage}[]{0.5\textwidth}
\centering
\includegraphics[width=2.5in]{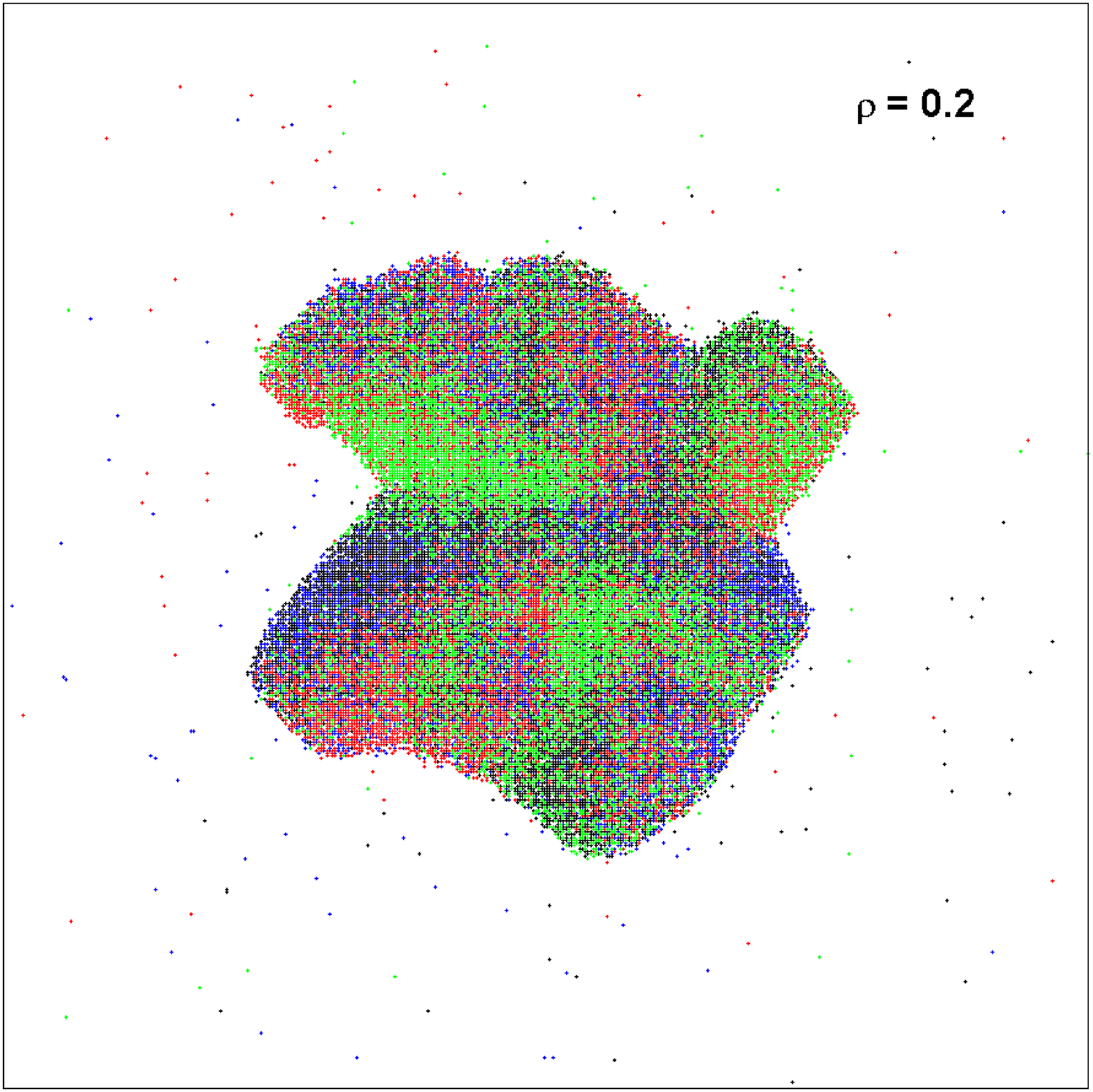}
\end{minipage}}%
\subfigure[]{
\label{j_2_10000:d}
\begin{minipage}[]{0.5\textwidth}
\centering
\includegraphics[width=2.5in]{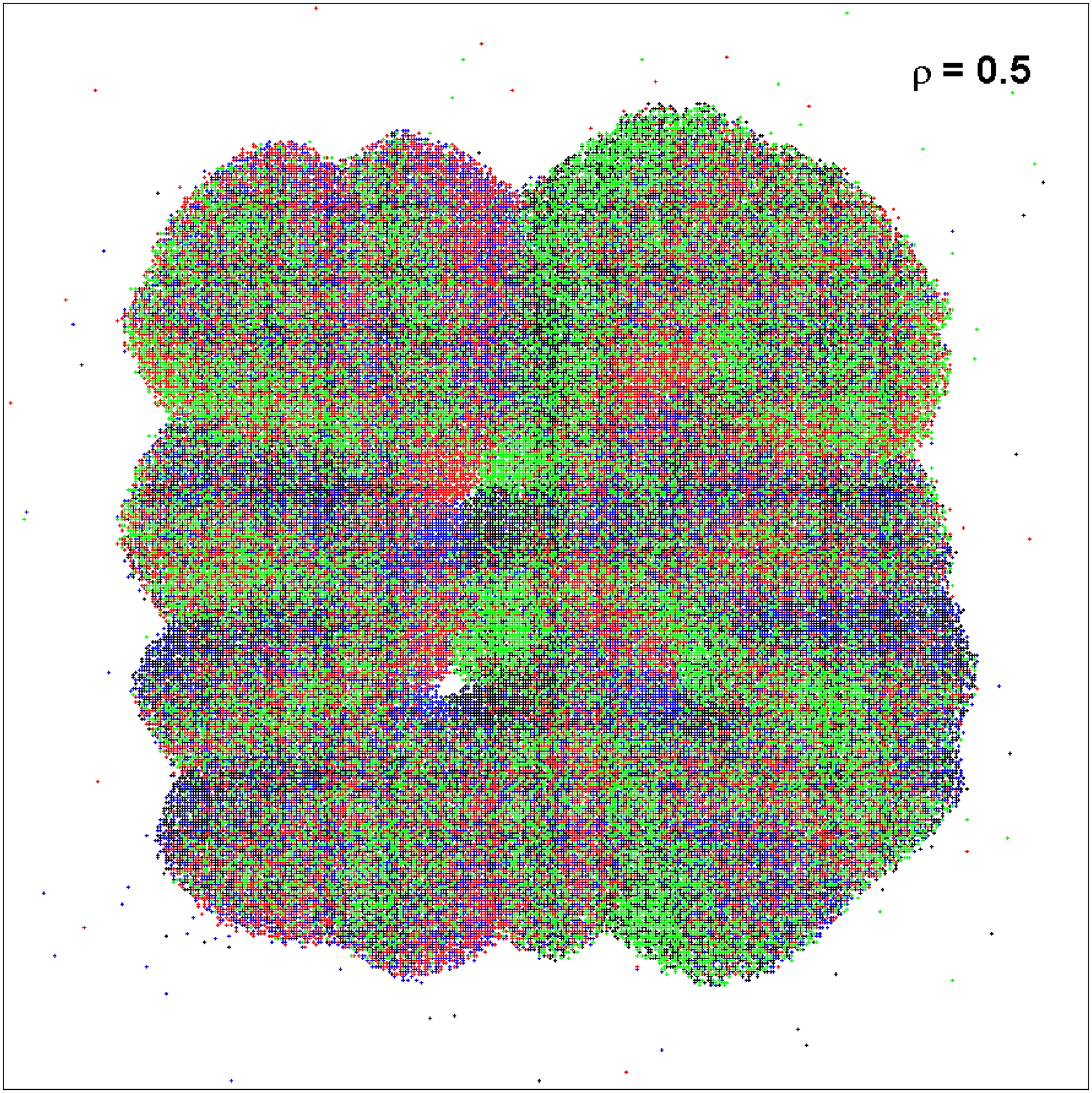}
\end{minipage}}
\caption{Model 2 - jam patterns at density $\rho=0.05, 0.07, 0.2, 0.5$ when system size $w=400$ and time step $t=10,000$. The objects whose destination points are in the upper left, upper right, lower left and lower right quadrants are colored with red, green, blue, and black colors, respectively.
}
\label{j_2_10000}
\end{figure}

\begin{figure}[]
\subfigure[]{
\label{j_3_10000:a}
\begin{minipage}[]{0.5\textwidth}
\centering
\includegraphics[width=2.5in]{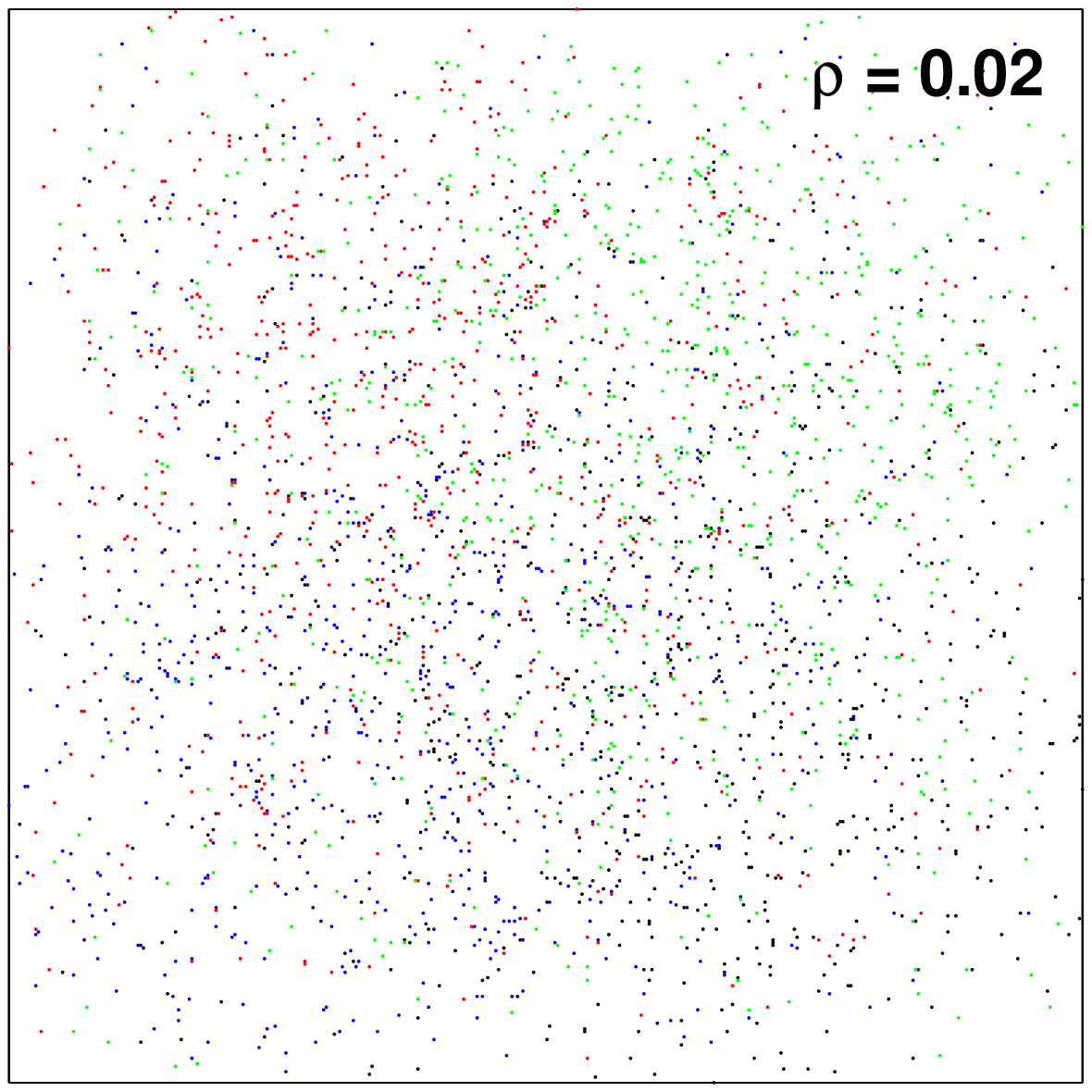}
\end{minipage}}%
\subfigure[]{
\label{j_3_10000:b}
\begin{minipage}[]{0.5\textwidth}
\centering
\includegraphics[width=2.5in]{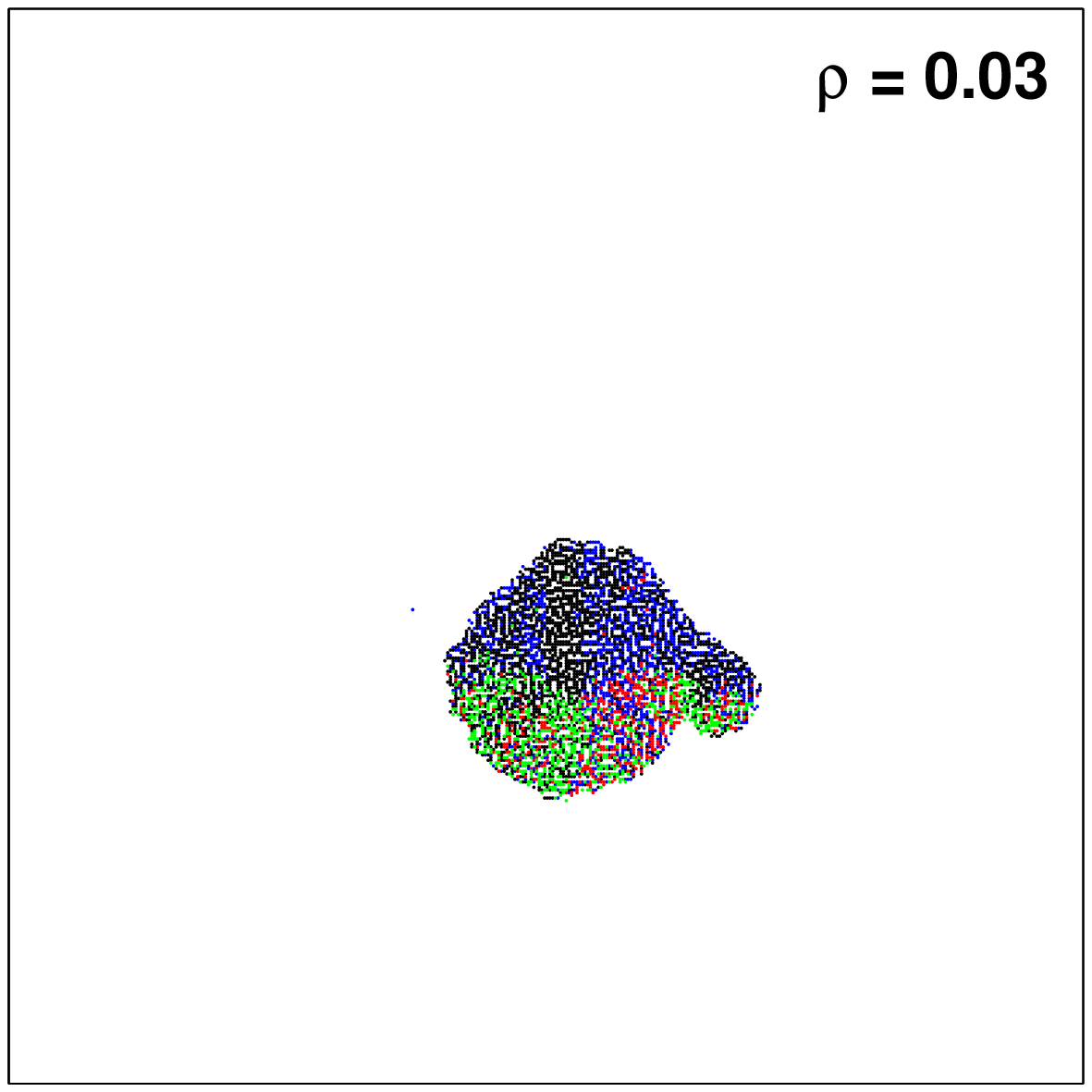}
\end{minipage}} \\
\subfigure[]{
\label{j_3_10000:c}
\begin{minipage}[]{0.5\textwidth}
\centering
\includegraphics[width=2.5in]{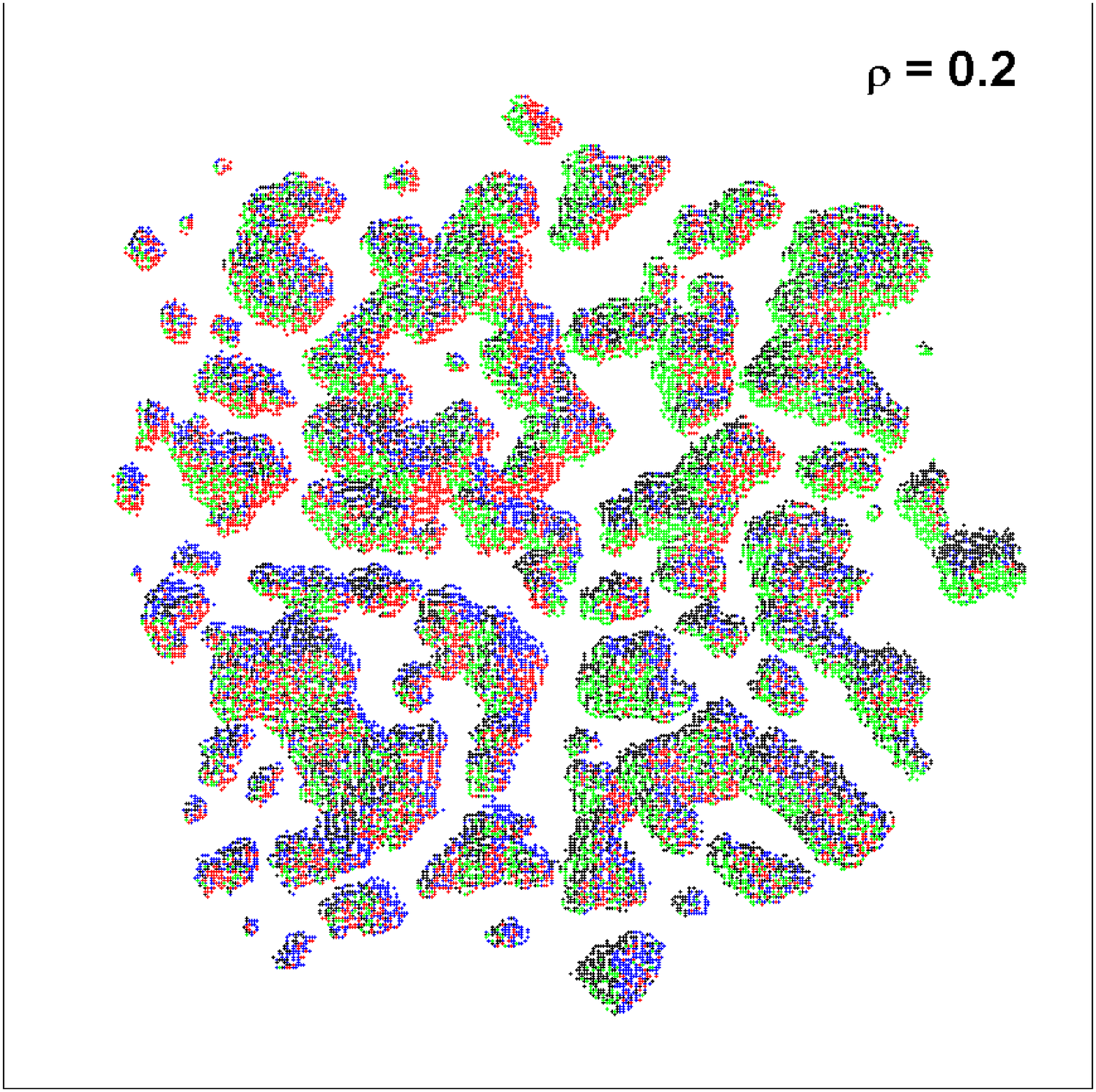}
\end{minipage}}%
\subfigure[]{
\label{j_3_10000:d}
\begin{minipage}[]{0.5\textwidth}
\centering
\includegraphics[width=2.5in]{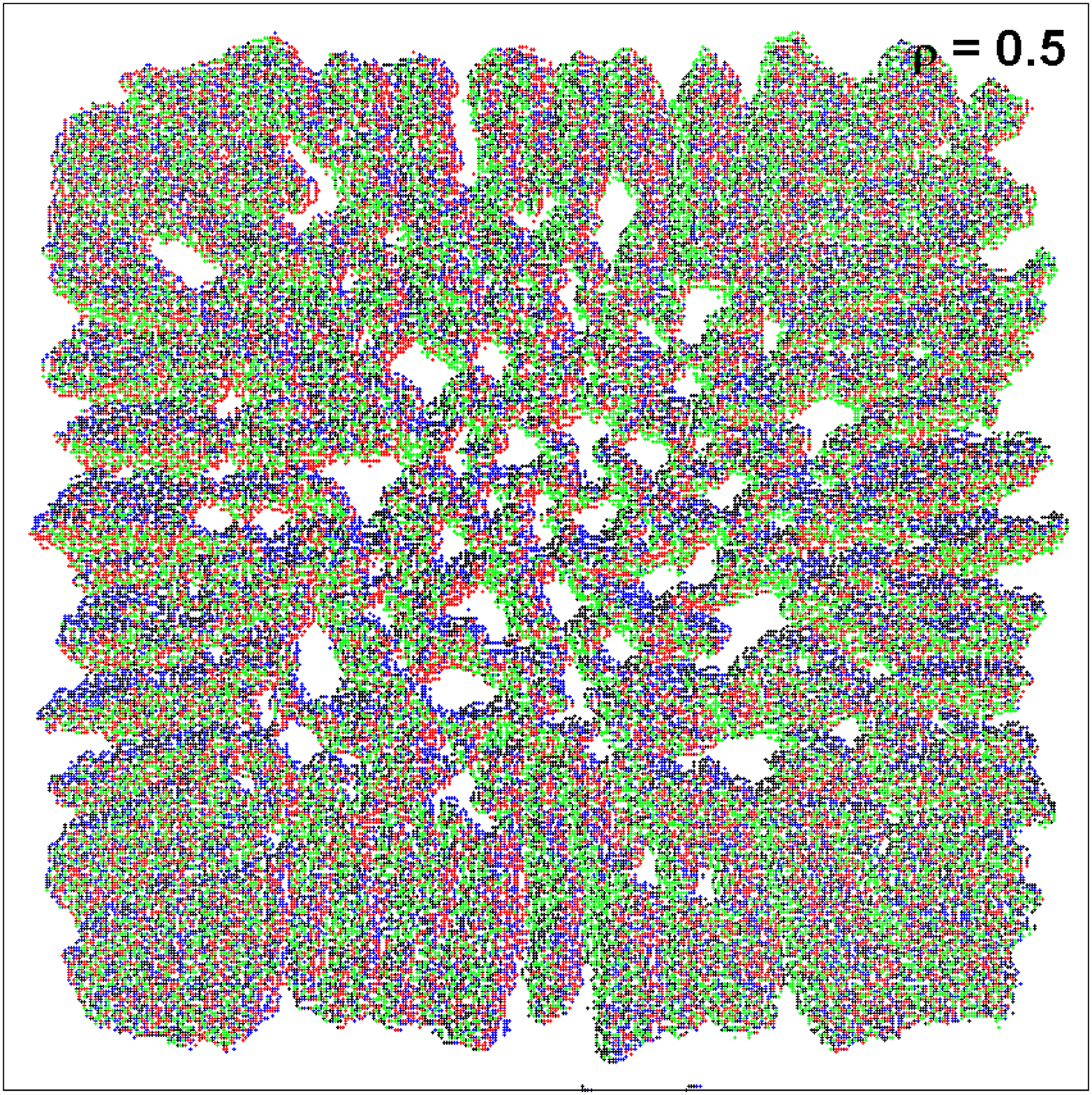}
\end{minipage}}
\caption{Model 3 - jam patterns at density $\rho=0.02, 0.03, 0.2, 0.5$ when system size $w=400$ and time step $t=10,000$. The objects whose destination points are in the upper left, upper right, lower left and lower right quadrants are colored with red, green, blue, and black colors, respectively.
}
\label{j_3_10000}
\end{figure}

We compare the jam patterns of model 2 with model 3 during the stationary stage for different density values. in Fig.~\ref{j_2_10000} and Fig.~\ref{j_3_10000}. In Fig.~\ref{j_2_10000},  the objects move freely at density 0.05 and get almost completely jammed at density 0.07, which coincides with the critical density in Fig.~\ref{v:b}. In Fig.~\ref{j_3_10000}, the objects move freely at density 0.02 and get into partial jam at density 0.03, which coincides with the critical density in Fig.~\ref{v:c}. For model 2 during jam phase, there is only one jam cluster in the middle of the lattice which gets larger with increasing the density. For model 3 during jam phase, however, the jam pattern gives a picture of ``islands group'', i.e. a group of little jam clusters scattering on the lattice. The number of clusters grows with increasing the density until the density has reached around 0.35. After that, these clusters gradually merge together to form a large cluster.

\begin{figure}[]
\centering
\subfigure[]{
\label{yanshi:a} 
\includegraphics[width=2.3in]{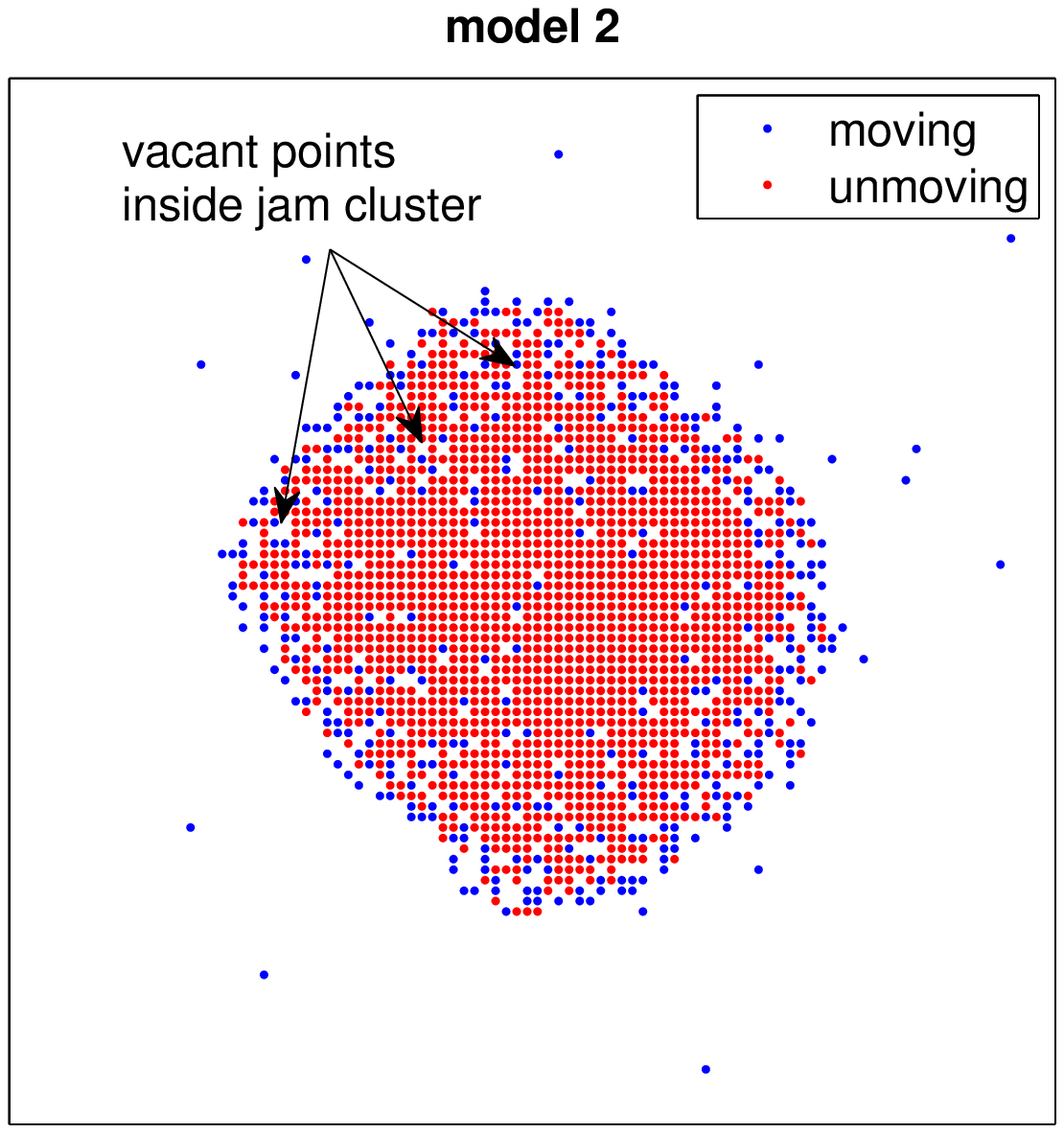}}
\subfigure[]{
\label{yanshi:b} 
\includegraphics[width=2.3in]{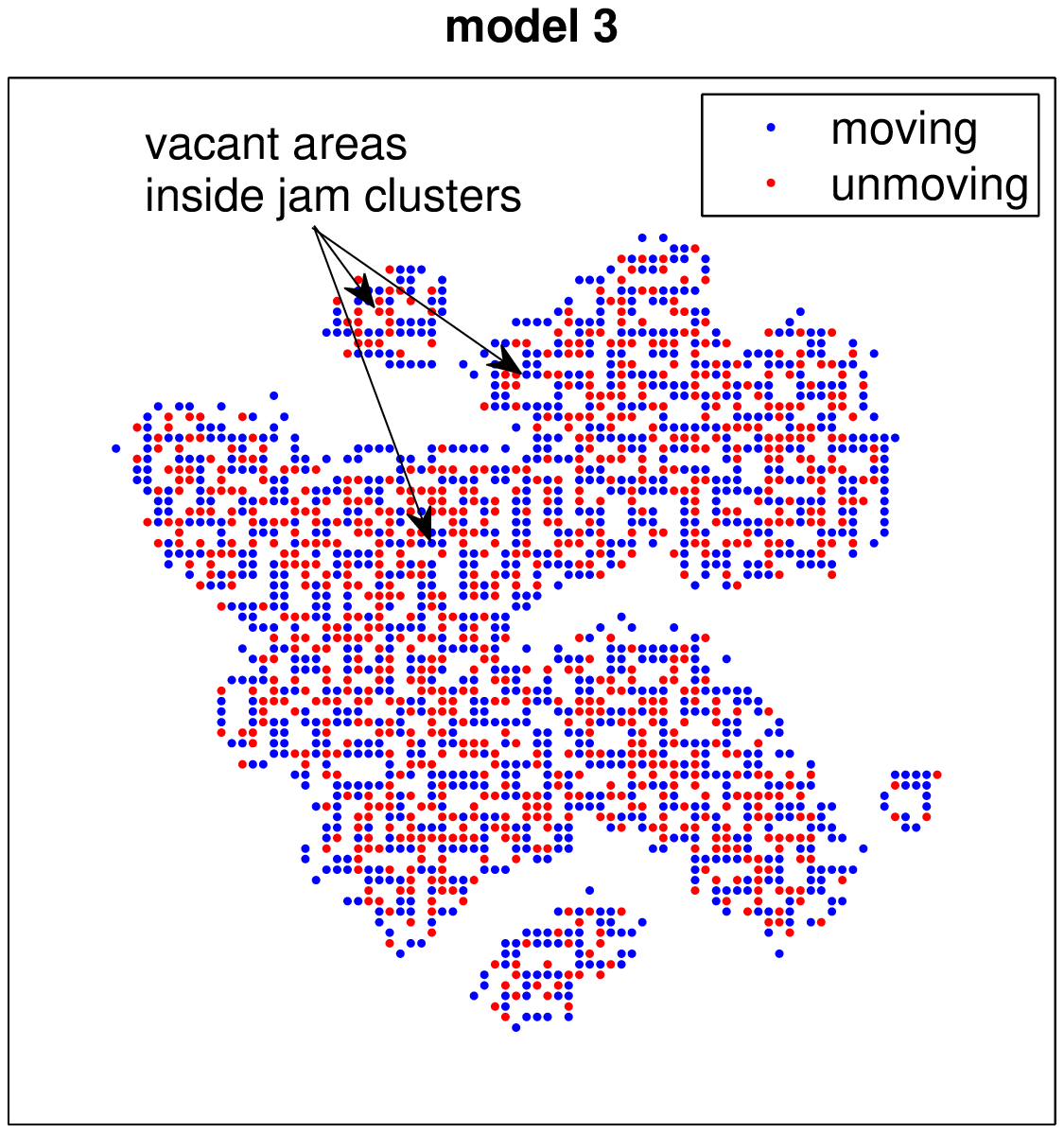}}
\caption{The moving objects and unmoving objects at time step $t=10001$ for model 2 and 3 when system size $w=100$ and density $\rho=0.2$.
}
\label{yanshi} 
\end{figure}

In Fig.~\ref{yanshi:a}, there are thinly scattered ``cavities'', or vacant points, across the jam cluster for model 2. The moving objects are unevenly distributed throughout the lattice. Most of moving objects lie along the surface of the cluster and the rest lie outside or in the middle of the cluster. This configuration can explain the low average velocity at jam phase in Fig.~\ref{v:a} and \ref{v:b}. Compared with model 2, there are dense vacant areas inside the jam cluster for model 3 shown in Figure.~\ref{yanshi:b}. The moving objects are evenly distributed at the edge of vacant areas inside the group of clusters. The number of moving objects is almost the same as that of unmoving objects. This configuration can explain the relatively high average velocity around 0.55 in Fig.~\ref{v:c} and \ref{v:d}. However, the objects in vacant areas are locked and they move only inside those vacant areas for model 3. The locked objects can't escape from jam clusters, even from their parent vacant areas. So an object can't arrive at its destination whenever it is trapped inside a vacant area in a jam cluster. This configuration can explain why the average flow of the system is so slow even when the average velocity reaches around 0.55 in Fig.~\ref{f:c} and \ref{f:d}.

\begin{figure}[]
\subfigure[]{
\label{c_t:a}
\begin{minipage}[]{0.5\textwidth}
\centering
\includegraphics[width=2.5in]{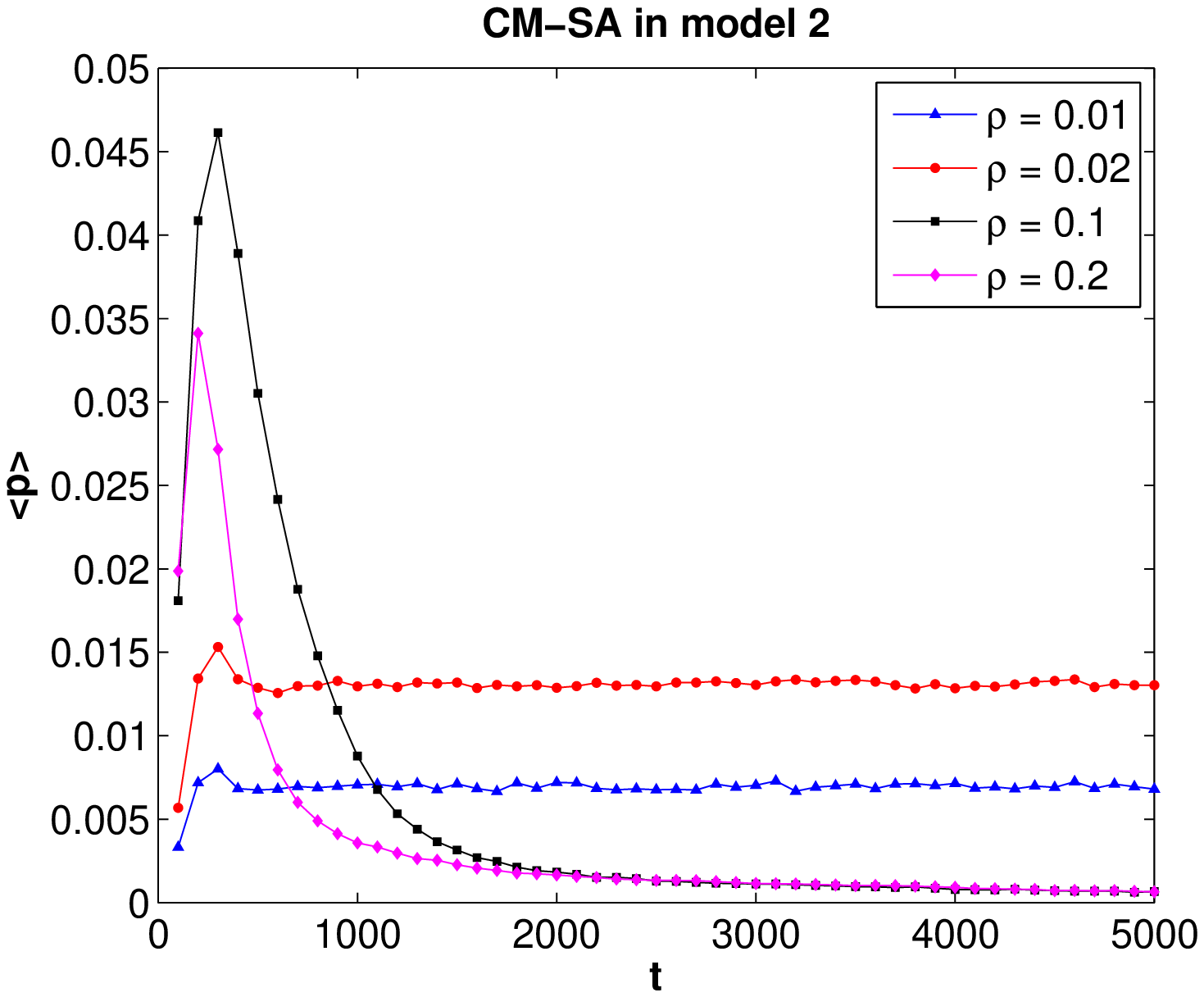}
\end{minipage}}%
\subfigure[]{
\label{c_t:b}
\begin{minipage}[]{0.5\textwidth}
\centering
\includegraphics[width=2.5in]{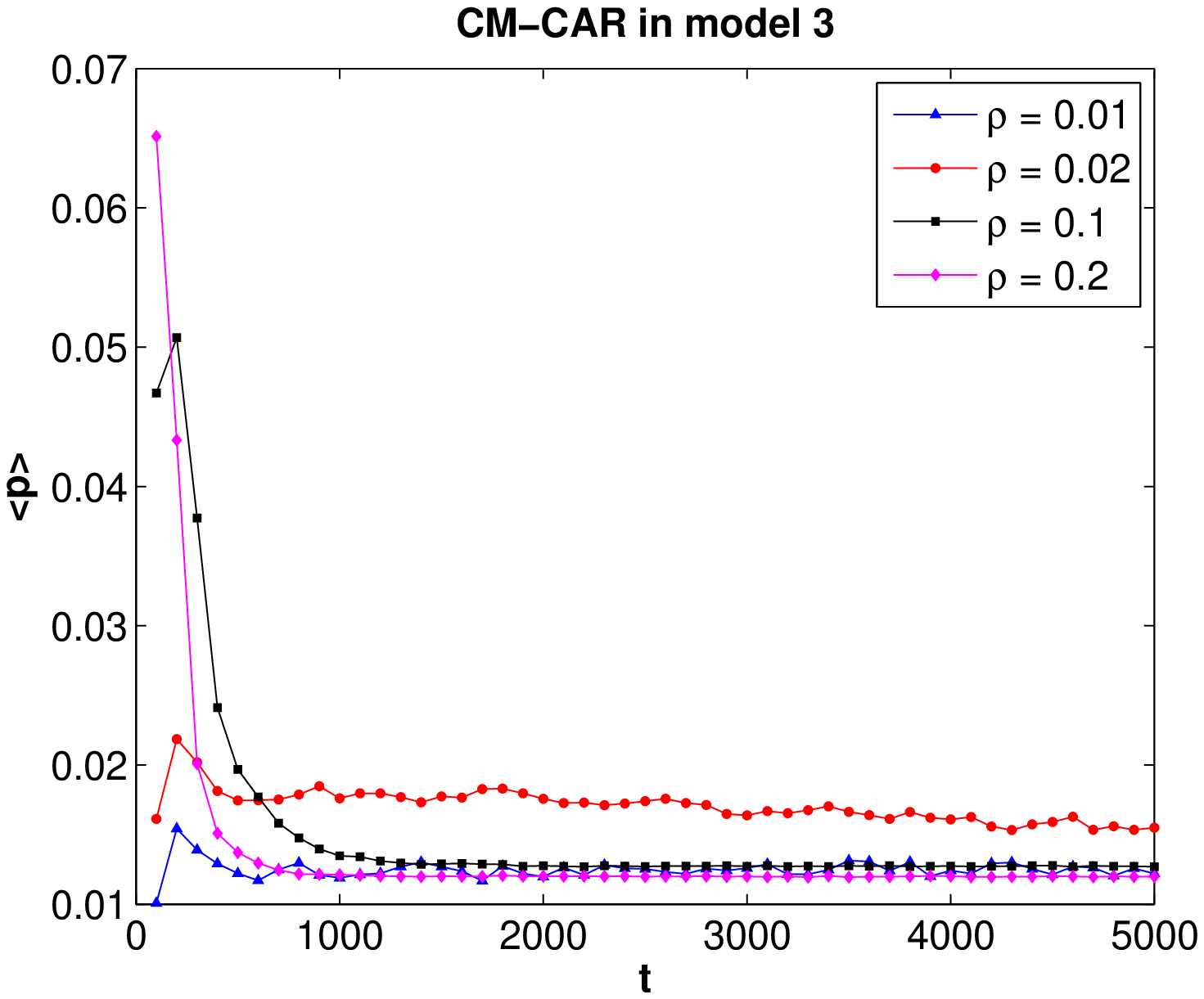}
\end{minipage}} \\
\subfigure[]{
\label{c_t:c}
\begin{minipage}[]{0.5\textwidth}
\centering
\includegraphics[width=2.5in]{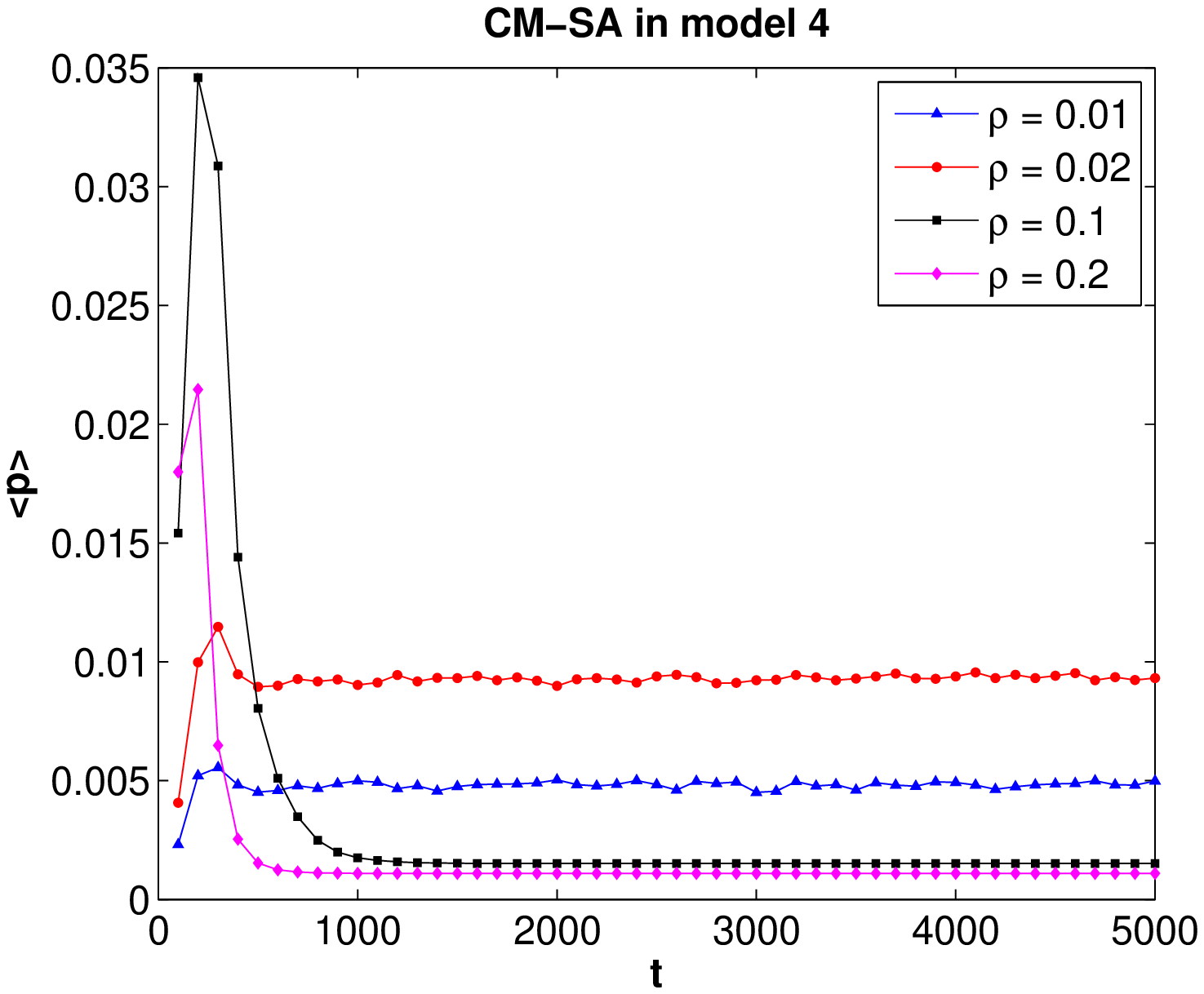}
\end{minipage}}%
\subfigure[]{
\label{c_t:d}
\begin{minipage}[]{0.5\textwidth}
\centering
\includegraphics[width=2.5in]{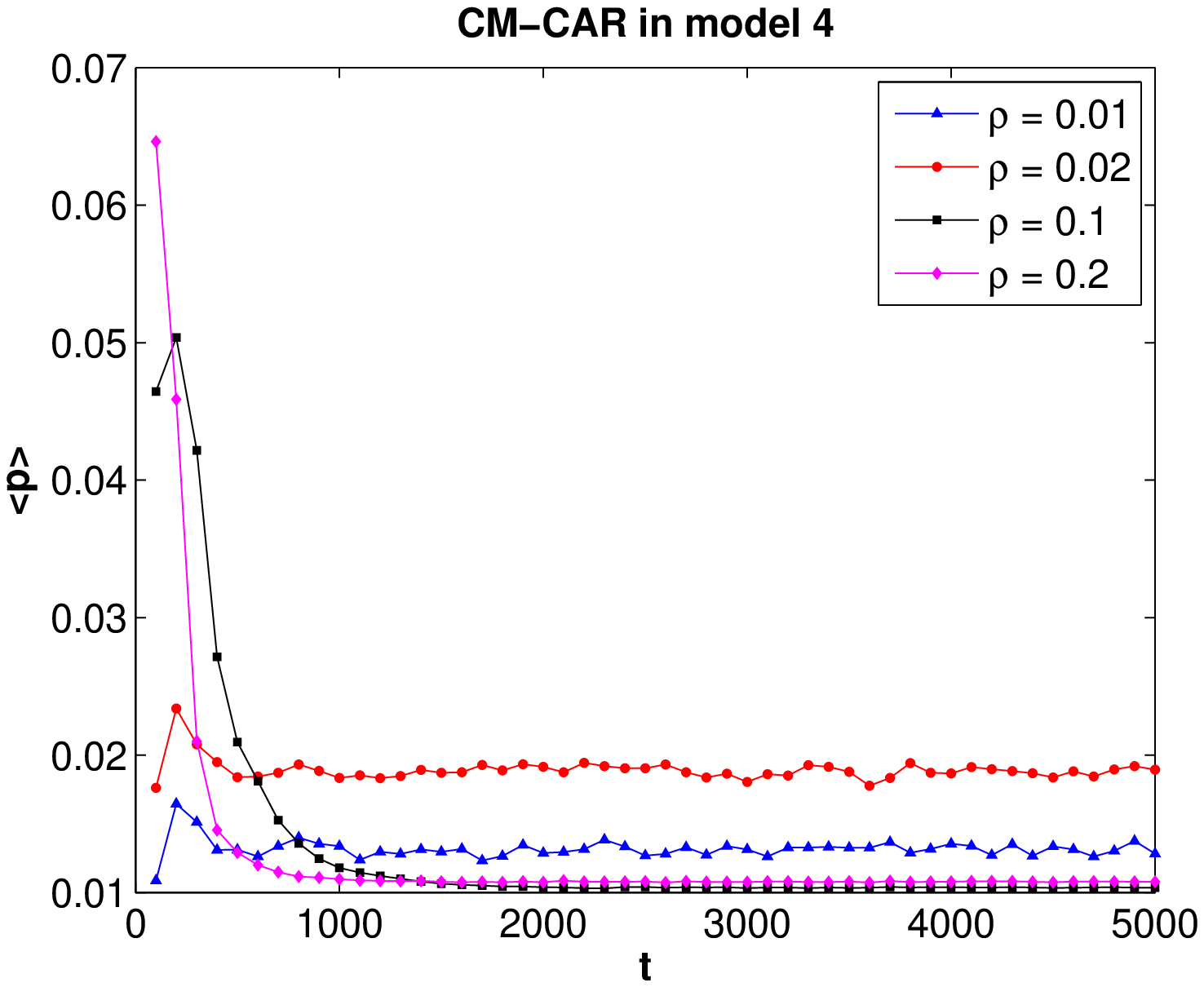}
\end{minipage}}
\caption{Plots of average cooperation probability $\langle p\rangle$ against time step $t$ for model 2, 3 and 4 when system size $w=400$ and density $\rho=0.01, 0.02, 0.1, 0.2$.
}
\label{c_t}
\end{figure}

In order to show the evolution of cooperation intensity, we investigate the average cooperation probability $\langle p\rangle$ of all objects as it changes through time. The cooperation probability $p$ of each object is defined as the average number of taking part in cooperation for each object during a period of time. The average cooperation probability $\langle p\rangle$ is defined as the average of $p$ over all objects at one time step. The CM-SA takes place when a pair of objects stands face-to-face and there is a vacant point on the side. The CM-CAR takes place when several objects compete for the same point and some of them have more than one candidate to move. The results are shown in Fig.~\ref{c_t}. During free phase, e.g. for $\rho=0.01, 0.02$, the shapes of probability curves are almost the same. The cooperation probability $\langle p\rangle$ firstly increases as time goes by and decreases after reaching the max value. After time step $t=500$ the value of $\langle p\rangle$ remains steady. The density is higher, the value of $\langle p\rangle$ is larger. It is noted that the cooperation probability of model 4 is lower than that of model 2. This is because the mechanism of choosing alternative routes incorporated by model 4 causes adjacent objects to bounce off each other and decreases the probability of remaining face-to-face.

\section{Summary and discussion} \label{conclusions}

The purpose of this paper is to study the jamming transition of two-dimensional point-to-point traffic through cooperative mechanisms. We proposed two cooperative mechanisms and incorporate them into the point-to-point traffic model: stepping aside (CM-SA) and choosing alternative routes (CM-CAR). Incorporating CM-SA is to prevent a type of ping-pong jumps from happening when two objects standing face-to-face want to move in opposite directions. Incorporating CM-CAR is to handle the conflict when more than one object competes for the same point.

It should be noted that the two cooperation mechanisms are both a type of decentralized cooperation mechanism (DCM). In this mechanism, several objects form a group temporarily to cooperate with each other. Usually, the number of group members is small and the group lasts for a short period. The group members take part in cooperation according to only their local environments. After this group splits up, the partnership of all members breaks up and they form some new groups with other objects again. Unlike the centralized cooperative mechanism (CCM), there is not a global mechanism, which can coordinate all objects or a major part of objects in the system.  The main advantages of DCM are due to its simple rules and requiring little or even no extra parameters. It is an interesting question to incorporate the CCM into our models, or combine the DCM with CCM in the same traffic model.

We investigate the four models mainly from fundamental diagrams, jam patterns and distribution of cooperation probability using computer simulation. It is found that the CM-SA used in model 2 increases the critical density and the average flow, although it decreases the average velocity a little. However, the CM-CAR used in model 3 and 4 doesn't realize our original intention in increasing the average flow. Although the average velocity is increased, the average flows for model 3 and 4 are lower than model 1 and 2. We investigate the jam patterns of four models carefully and explain this result qualitatively.

In addition, there are a number of remaining questions that are worth further research. Firstly, we are unclear whether the critical density remains non-zero in the thermodynamic limit. We have only carried out simulations on the lattice of size $w=400$ at most. Using larger regions is very costly in terms of computer simulation time. Secondly, the two cooperation mechanisms proposed in this paper are both a type of decentralized cooperation. Incorporating the centralized cooperative mechanisms, even combining them with decentralized cooperative mechanisms, is an interesting and useful question.


\begin{thebibliography}{00} 
\bibitem{Chowdhury2000} D. Chowdhury, L. Santen, A. Schadschneider, Phys. Rep. 329 (2000) 199.
\bibitem{Helbing2001} D. Helbing, Rev. Mod. Phys. 73 (2001) 1067.
\bibitem{Nagatani2002} T. Nagatani, Rep. Prog. Phys. 65 (2002) 1331.
\bibitem{Biham1992} O. Biham, A. A. Middleton, D. Levine, Phys. Rev. A 46 (1992) 6124.
\bibitem{Nagatani1993} T. Nagatani, J. Phys. Soc. Jap. 62 (1993) 2656.
\bibitem{Nagatani1995} T. Nagatani, J. Phys. Soc. Jap. 64 (1995) 1421.
\bibitem{Torok1996} J. T\"{o}r\"{o}k, J. Kert\'{e}sz, Physica A 231 (1996) 515.
\bibitem{Nagatani2009a} T. Nagatani, Phys. Lett. A 373 (2009) 536. 
\bibitem{Benyoussef2003} A. Benyoussef, H. Chakib, H. Ez-Zahraouy, Phys. Rev. E 68 (2003) 026129.
\bibitem{Nagatani2008} T. Nagatani, Phys. Lett. A 372 (2008) 5887. 

\bibitem{Nagatani2009b} T. Nagatani, Physica A 388 (2009) 14.
\bibitem{Chowdhury1999} D. Chowdhury, A. Schadschneider, Phys. Rev. E 59 (1999) R1311.
\bibitem{Brockfeld2001} E. Brockfeld, R. Barlovi\'{c}, A. Schadschneider, M. Schreckenberg, Phys. Rev. E 64 (2001) 056132.
\bibitem{Shi2007} X. Shi, Y. Wu, H. Li, R. Zhong, Physica A 385 (2007) 659.
\bibitem{Moussa2007} N. Moussa, Int. J. Mod. Phys. C 18 (2007) 1047.
\bibitem{In-nami2007} J. In-nami, H. Toyoki, Physica A 378 (2007) 485.
\bibitem{Huang2007} D. Huang, W. Huang, Chin. J. Phys. 45 (2007) 708.
\bibitem{Fang2010} J. Fang, J. Shi, X.-Q. Chen, Z. Qin, Int. J. Mod. Phys. C 21 (2010) 221. 

\bibitem{Muramatsu1999} M. Muramatsu, T. Irie, T. Nagatani, Physica A 267 (1999) 487. 
\bibitem{Muramatsu2000a} M. Muramatsu, T. Nagatani, Physica A 275 (2000) 281. 
\bibitem{Muramatsu2000b} M. Muramatsu, T. Nagatani, Physica A 286 (2000) 377. 
\bibitem{Jiang2007} R. Jiang, Q.-S. Wu, Physica A 373 (2007) 683.
\bibitem{Ohira1998} T. Ohira, R. Sawatari, Phys. Rev. E 58 (1998) 193.
\bibitem{Tretyakov1998} A. Tretyakov, H. Takayasu, M. Takayasu, Physica A 253 (1998) 315.
\bibitem{Beckers1992} R. Beckers, J. L. Deneubourg, S. Goss, J. Theor. Biol. 159 (1992) 397.
\bibitem{Chowdhury2002} D. Chowdhury, V. Guttal, K. Nishinari, A. Schadschneider, J. Phys. A 35 (2002) L573.
\bibitem{John2004} A. John, A. Schadschneider, D. Chowdhury, K. Nishinari, J. Theor. Biol. 231 (2004) 279.
\bibitem{Mitchell1996} M. Mitchell, B. Durnota, Complexity International 3 Paper ID: mitchdur (1996).
\bibitem{Moussa2005} N. Moussa, Int. J. Mod. Phys. C 16 (2005) 1849. 
\bibitem{Maniccam2004} S. Maniccam, Physica A 331 (2004) 669.
\bibitem{Maniccam2005} S. Maniccam, Physica A 346 (2005) 631. 
\bibitem{Maniccam2006} S. Maniccam, Physica A 363 (2006) 512. 

\bibitem{Burstedde2001} C. Burstedde, K. Klauck, A. Schadschneider, J. Zittartz, Physica A 295 (2001) 507.
\bibitem{Kirchner2004} A. Kirchner, H. Kl\"{u}pfel, K. Nishinari, A. Schadschneider, M. Schreckenberg, J. Stat. Mech.-Theory Exp. (2004) P10011.

\end{thebibliography}
\end{document}